\documentclass[12pt]{article}
\usepackage[utf8]{inputenc}
\usepackage[russian,ukrainian]{babel}
\usepackage[OT1]{fontenc}
\usepackage{amsmath}
\usepackage{amsfonts}
\usepackage{amssymb}
\usepackage{graphicx}
\usepackage[left=2cm,right=2cm,top=2cm,bottom=2cm]{geometry}
\usepackage[colorlinks=true,citecolor=blue,filecolor=magenta,unicode]{hyperref}
\usepackage[numbers,sort&compress]{natbib}
\newcommand*{\doi}[1]{\href{https://doi.org/#1}{doi: #1}}

\title{Метод Лапласа для найпростіших діаграм пружного розсіяння скалярних частинок}

\author{І.В. Шарф, Т.М. Зеленцова, Н.О. Чудак, О.С. Потієнко, Д.А. Пташинський, \\
	 К.К. Меркотан, Т.В. Юшкевич, А.О. Мілєва}

\begin{document}
	
\maketitle
	
	\begin{center}
		\begin{minipage}{15cm}
 \textit{Одеський національний політехнічний університет, \\
	\mbox{~~}Проспект Шевченка 1, Одеса, 65044, Україна} \\
				
			\end{minipage}
		\end{center}	
	
\begin{abstract}
Запропоновано алгоритм застосування методу Лапласа для розрахунку найпростішої діаграми Р. Фейнмана з однією петлею, що виникає в найпростішій скалярній теорії ${{\phi }^{3}}$. Розрахунок внеску такої діаграми в амплітуду розсіяння потребує обчислення чотирикратного інтегралу по компонентах чотири імпульсу, що циркулює по петлі. Суть методу Лапласа для розрахунку багатократних інтегралів полягає в тому, що якщо модуль підінтегрального виразу має всередині області інтегрування точку достатньо гострого максимуму, то інтеграл можна замінити гаусовим інтегралом шляхом представлення підінтегральної функції в виді експоненти від її логарифму і розкладу цього логарифму в ряд Тейлора в околі точки максимуму з обмеженням доданками другого ступеню. В роботі показано, що всередині чотиривимірної області інтегрування є дві двовимірні поверхні, які не перетинаються між собою, і на яких досягається максимум модуля підінтегрального виразу. При цьому виникає проблема, яка полягає в тому, що підінтегральний вираз неаналітично залежить від параметрів, які відповідають за обхід полюсів. Також неаналітично від цих параметрів залежать і похідні від логарифму амплітуди розсіяння. Проте, в роботі показано, що ці неаналітичності компенсують одна одну. Так відбувається тому, що при наближенні параметрів, які визначають обхід полюсів, до нуля, неаналітичність амплітуди спрямовує «висоту» максимуму до нескінченості, а неаналітичність похідних наближує «ширину» максимуму до нуля. Оскільки основний внесок в інтеграл робить малий окіл кожної з поверхонь, то в роботі проведено розгляд внеску кожного з цих околів окремо, а внесок діаграми в амплітуду розсіяння розраховано як суму внесків цих околів. Для кожного з цих внесків шляхом переходу до відповідних змінних інтегрування вдається виключити неаналітичність, після чого параметри, які відповідають за обхід полюсів можуть бути наближені до нуля. В результаті такої процедури три з чотирьох інтегрувань можна виконати аналітично і розрахунок внеску діаграми в амплітуду розсіяння зводиться до чисельного обчислення однократного інтегралу в скінчених границях від виразу, який не містить неаналітичностей. Описаний метод розрахунку застосовано до побудови моделі залежності диференційного перерізу пружного розсіяння  $\frac{d{{\sigma }_{el}}}{dt}\left( t \right)$ від квадрату переданого чотириімпульсу $ t $– манделстамівської змінної. Розрахунок проводився до діаграм, отриманих в межах методу багаточастинкових полів. Отриманий результат призводить до скінченого значення амплітуди розсіяння при $ t=0 $ (на відміну від теорії збурень КХД).	
	
\end{abstract}

\section{Вступ}
\begin{figure}
	\centering
	\includegraphics[width=0.7\linewidth]{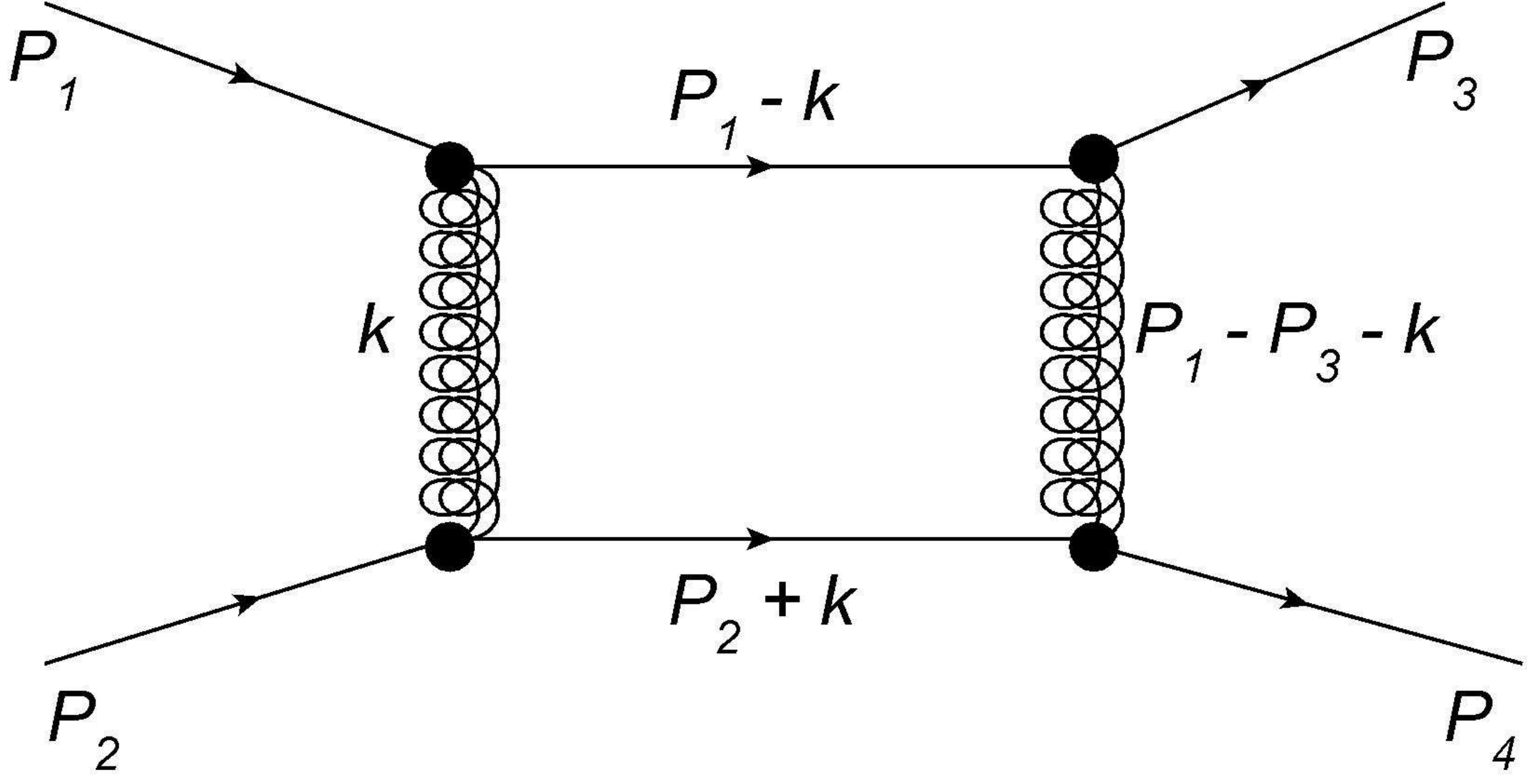}
	\caption[]{Найпростіша петльова діаграма пружного розсіяння мезонів. Горизонтальні лінії відповідають розповсюдженню мезонів, де ${{P}_{1}},{{P}_{2}}$ - чотириімпульси вхідних мезонів, ${{P}_{3}},{{P}_{4}}$ - чотириімпульси вихідних мезонів, $k,{{P}_{1}}-k,{{P}_{2}}+k,{{P}_{1}}-{{P}_{3}}-k$ - чотириімпульси віртуальних частинок. Подвійні лінії відповідають зв’язаному стану глюонів – глюболам.}
	\label{fig:naiprostihadiagrama1}
\end{figure}

На протязі довгого часу діаграми Фейнмана використовуються як інструмент для розрахунку спостережних в експерименті квантово-польових процесів \cite{PhysRev.76.769}, проте достатньо ефективного методу розрахунку внесків діаграм з петлями на сьогоднішній день не існує. Труднощі розрахунків інтегралів, що відповідають діаграмам Фейнмана разом із сумнівами в можливості застосування теорії збурень у випадку великих значень константи зв'язку призвели до формулювання різних феноменологічних підходів до розрахунків експериментальних характеристик розсіяння \cite{Pancheri2017,Dryomin:2013, book:1030746, Gribov:1968fc, BAKER19761, Kaidalov:2003, Bourrely:2002wr, FADIN197550, Lipatov:2008, Kuraev:1977fs}, що дозволяють уникнути прямих розрахунків багатовимірних інтегралів по чотириімпульсах віртуальних частинок. Серед цих моделей, найбільш популярними є ті, що пов'язані з реджіонною теорією. Проблеми цих моделей детально розглядалися в роботах \cite{Sharf:2011ufj, Sharph:2011wm}.
Ще одним напрямком подолання труднощів, пов'язаних із розрахунком інтегралів, що відповідають багатопетльовим діаграмам Фейнмана є застосування методу Монте-Карло для багатовимірних інтегралів \cite{PhysRevD.61.125001,doi:10.1142/S0217751X08040263,Li_2016}. Але застосування методу Монте-Карло є складною процедурою, яка потребує великого обсягу чисельних розрахунків і при цьому є малоефективною з точки зору виявлення фізичних механізмів, відповідальних за ту чи іншу поведінку величин, що спостерігаються в експерименті. Метод, який ми збираємось запропонувати в цій роботі є значно простішим і потребує значно меншого об'єму обчислень. Цей метод оснований на застосуванні відомого методу Лапласа \cite{book:5160} для багатовимірних інтегралів.
Застосування методу Лапласа значно спрощує розрахунки внесків у перерізи безпетльових, а також деяких петльових діаграм Фейнмана \cite{Sharf:2011ufj,Sharph:2011wm,Sharf:2012vy}. Всі діаграми, що розглядалися в згаданих роботах мали таку властивість, що особливі точки підінтегральних виразів, до яких застосовувався метод Лапласа, знаходилися поза областю інтегрування. Це дозволяло малі комплексні параметри фейнманівських знаменників, які відповідають за «правильний» обхід полюсів (ми будемо позначати їх далі як зазвичай $ i\varepsilon  $), наблизити до нуля ще до розрахунку інтегралу.
При розгляді діаграм пружного розсіяння, ми не маємо такої спрощеної обставини. Для діаграм пружного розсіяння в області інтегрування можна виділити такі її підмножини, на яких підінтегральний вираз стане нескінченим і інтеграл втратить сенс, якщо згадані малі параметри наблизити до нуля. В такому випадку наблизити параметри до нуля ще до розрахунку інтегралу неможливо. Тому до проблеми чисельного розрахунку інтегралу додається проблема чисельного здійснення граничного переходу. З іншого боку, оскільки інтеграл в кінцевому рахунку існує, це означає, що нескінченості підінтегрального виразу, що виникають при наближенні  $ i\varepsilon  $ до нуля повинні компенсуватися наближенням до нуля мір цих підмножин, на яких ці нескінченості досягаються. Тобто при ненульовому, але малому $ i\varepsilon  $ до граничного переходу підінтегральний вираз буде мати різкий максимум, «висота» якого наближатиметься до нескінченості при  $ i\varepsilon \to 0.$ Але й «ширина» максимуму також наближатиметься до нуля при $ i\varepsilon \to 0$ і ці два ефекти повинні компенсувати один одного. Вже ця обставина могла б дозволити застосувати до розрахунку інтегралу метод Лапласа, в якому роль «ширини» максимуму визначається детермінантом матриці других похідних від логарифму підінтегрального виразу в околі точки максимуму. Однак метод Лапласа можна застосувати простіше і математично більш обґрунтовано за допомогою алгоритму, який пояснимо на прикладі найпростіших петльових діаграм пружного розсіяння. Цей алгоритм планується застосовувати для діаграм пружного розсіяння з довільною кількістю петель. 
Метою роботи є запропонувати ефективний метод розрахунку діаграм Р.Фейнмана і застосувати його на прикладі найпростішої діаграми з однією петлею.
Розглянемо найпростішу діаграму пружного розсіяння з однією петлею (Рис.\ref{fig:naiprostihadiagrama1}). Оскільки у експериментах з пружного розсіяння адронів, наприклад протон-протонного розсіяння, найбільший інтерес представляє модель взаємодіючих багаточастинкових \cite{Chudak:2016,articleJFS}, біспінорного (протонного) і глюбольного полів. Однак така модель потребує великого об’єму обчислень, тому в даній роботі ми пропонуємо розглянути модель із скалярними частинками, яка відповідає розсіянню мезонів.
Аналітичний вираз, що відповідає діаграмі на Рис.\ref{fig:naiprostihadiagrama1} має вид:

\begin{equation}\label{fi3_amplituda}
\begin{split}
&A_1  = G^4 \int {d^4 k} \frac{1}{{M_\mu ^2  - \left( {P_1  - k} \right)^2  - i\varepsilon }}\frac{1}{{M_\mu ^2  - \left( {P_2  + k} \right)^2  - i\varepsilon }} \times  \\ 
&\times \frac{1}{{M_G^2  - k^2  - i\varepsilon }}\frac{1}{{M_G^2  - \left( {P_1  - P_3  - k} \right)^2  - i\varepsilon }}. \\ 
\end{split}
\end{equation}
Позначення ${{M}_{\mu }}$ i ${{M}_{G}}$ означають масу мезону і масу глюболу відповідно. Всі величини розглядаються обзрозміреними на масу мезону. Позначення $ G $ означає  ефективну константу зв'язку. Маса глюболу і константа  $ G $ озглядаються як підгінні параметри задачі.

Якщо ми за допомогою методу Лапласа будемо шукати точки максимуму підінтегрального виразу до $ i\varepsilon \to 0$, то значення підінтегральної функції в точці максимуму буде наближатися до нескінченності. Окрім того, сама функція, а відтак i її логарифм не можна буде представити рядом Тейлора і обмежитися в цьому ряді доданками другого порядку. Для розв’язку зазначеної проблеми ми пропонуємо наступний підхід. Розглянемо ситуацію, коли $ i\varepsilon$  ще не наближене до 0, але має мале значення. Підінтегральний вираз в формулі \eqref{fi3_amplituda} містить добуток чотирьох знаменників. Під «знаменниками» будемо мати на увазі вирази:

\begin{equation}\label{znamennici}
\begin{array}{l}
Z_1  = M_\mu ^2  - \left( {P_1  - k} \right)^2 ,Z_2  = M_\mu ^2  - \left( {P_2  + k} \right)^2 , \\ 
Z_3  = M_G^2  - k^2 ,Z_4  = M_G^2  - \left( {P_1  - P_3  - k} \right)^2 . \\ 
\end{array}
\end{equation}

Тобто, мова йде про вирази, які залишилися б у знаменниках дробів у \eqref{fi3_amplituda} після граничного переходу  $ i\varepsilon \to 0$ . Внаслідок малості основний внесок в інтеграл буде вносити область, в якій найбільша кількість знаменників \eqref{znamennici} дорівнює нулю. Якщо цю кількість позначити як $n$, то на ній підінтегральний вираз матиме величину ${1 \mathord{\left/{\vphantom {1 {\varepsilon ^n }}} \right.	\kern-\nulldelimiterspace} {\varepsilon ^n }}$  і чим більше буде $n$, тим більшою буде ця величина. 
  
\section{Системи рівнянь для одночасного обернення в нуль знаменників \eqref{znamennici}}
   
Оскільки величина $ {{A}_{1}}, $ яка визначається співвідношенням \eqref{fi3_amplituda} є Лоренц-інваріантною, її можна розраховувати в будь-якій системі відліку. З міркувань зручності розрахунку в якості системи відліку оберемо систему центру мас вихідних частинок. При цьому третю координатну вісь системи координат спрямуємо уздовж тривимірного імпульсу ${{\vec{P}}_{1}},$  першу оберемо перпендикулярної третій, але так щоб вона належала площині імпульсів ${{\vec{P}}_{1}}$ і ${{\vec{P}}_{3}}$, а другу - перпендикулярно площині першої і третьої так, щоб утворилася права система координат.
   
Розглянемо системи рівнянь, які утворюватимуться, якщо одночасно прирівняти до нуля два знаменники з тих, що наведені в \eqref{znamennici}. Розглянемо спочатку такі комбінації, в яких один із знаменників відповідає горизонтальній лінії на рис.\ref{fig:naiprostihadiagrama1}, а один - вертикальній. Почнемо з пари знаменників $Z_2 $ і $Z_3$. Якщо записати для цих знаменнників вирази по відношенню до системи центру мас вихідних частинок, то отримаємо систему рівнянь
 
\begin{equation}\label{Z2Z3}
\begin{split}
\left\{ \begin{array}{l}
M_\mu^2  - \left( {{{\sqrt s } \mathord{\left/
			{\vphantom {{\sqrt s } 2}} \right.
			\kern-\nulldelimiterspace} 2} + k^0 } \right)^2  + \left( {k^1 } \right)^2  + \left( {k^2 } \right)^2  + \left( { - P + k^3 } \right)^2  = 0, \\ 
M_G^2  - \left( {k^0 } \right)^2  + \left( {k^1 } \right)^2  + \left( {k^2 } \right)^2  + \left( {k^3 } \right)^2  = 0. \\ 
\end{array} \right.
\end{split}
\end{equation}
Оскільки обидва вирази в лівіх частинах рівнянь системи \eqref{Z2Z3} повинні дорівнювати нулю, як наслідок - вони повинні дорівнювати один одному:
\begin{equation}\label{Z2Z3naslidok}
\begin{array}{l}
M_\mu^2  - \left( {{{\sqrt s } \mathord{\left/
			{\vphantom {{\sqrt s } 2}} \right.
			\kern-\nulldelimiterspace} 2} + k^0 } \right)^2  + \left( {k^1 } \right)^2  + \left( {k^2 } \right)^2  + \left( {P - k^3 } \right)^2  =  \\ 
= M_G^2  - \left( {k^0 } \right)^2  + \left( {k^1 } \right)^2  + \left( {k^2 } \right)^2  + \left( {k^3 } \right)^2 . \\ 
\end{array}
\end{equation}
З цього рівняння видно, що квадратичні по компонентам чотириветора $  k $ доданки однаково входять в обидві частини рівності. Тому іх можна відняти від обох частин рівності. В результаті отримаємо простий зв'язок між компонентами $k^0$ i $k^3$:
\begin{equation}\label{k0cherezk3}
k^0  =  - \frac{{2P}}{{\sqrt s }}k^3  - \frac{{M_G^2 }}{{\sqrt s }}.
\end{equation} 
Підставляючи \eqref{k0cherezk3} в друге рівняння системи \eqref{Z2Z3}, отримаємо:
\begin{equation}\label{Z2Z3rivnanna}
M_G^2  - \left( { - \frac{{2P}}{{\sqrt s }}k^3  - \frac{{M_G^2 }}{{\sqrt s }}} \right)^2  + \left( {k^1 } \right)^2  + \left( {k^2 } \right)^2  + \left( {k^3 } \right)^2  = 0.
\end{equation}
Розкриваючи в цьому виразі дужки і виділяючи повний квадрат по $ k^3 $ після перетворення отримаємо
\begin{equation}\label{ZaperechennaZ2Z3}
M_G^2 \left( {1 - \frac{{M_G^2 }}{{4M_\mu^2 }}} \right) + \left( {k^1 } \right)^2  + \left( {k^2 } \right)^2  + \frac{{4M_\mu^2 }}{s}\left( {k^3  - \frac{{M_G^2 P}}{{2M_\mu^2 }}} \right)^2  = 0.
\end{equation}
Приймаючи умову 
\begin{equation}\label{Umova}
M_G  \le 2M_\mu
\end{equation}
як це обговорювалося у вступі, бачимо що у випадку коли реалізується знак <<менше>> рівняння \eqref{ZaperechennaZ2Z3} не має жодного розв'язку, а у випадку  <<дорівнює>> будемо мати лише одну трійку чисел $k^1 ,k^2 ,k^3 $ яка задовольняє цьому рівнянню, але з одним значенням $ k^0 ,$ що знайдеться з \eqref{k0cherezk3} утворить множину нульової міри в області інтегрування \eqref{fi3_amplituda} і тому не дасть внеску в інтеграл. Отже рівняння системи \eqref{Z2Z3} є несумісними і знаменники $Z_2$ і $Z_3$ одночасно в нуль не обертаються.

Для першого і третього знаменників отримуємо систему
\begin{equation}\label{Z1Z3}
\left\{ \begin{array}{l}
M_\mu^2  - \left( {{{\sqrt s } \mathord{\left/
			{\vphantom {{\sqrt s } 2}} \right.
			\kern-\nulldelimiterspace} 2} - k^0 } \right)^2  + \left( {k^1 } \right)^2  + \left( {k^2 } \right)^2  + \left( {P - k^3 } \right)^2  = 0, \\ 
M_G^2  - \left( {k^0 } \right)^2  + \left( {k^1 } \right)^2  + \left( {k^2 } \right)^2  + \left( {k^3 } \right)^2  = 0. \\ 
\end{array} \right.
\end{equation}
Ця система відрізняється від \eqref{Z2Z3} лише заміною $ k^0  \to - k^0 .$ Роблячи таку заміну отримаємо ту ж систему що й  \eqref{Z2Z3}, відповідно з тим самим висновком, що така система не має розв'язків. Отже перший і третій знаменники одночасно нулю дорівнювати не можуть.  

Для першого і четвертого знаменників отримаємо систему
\begin{equation}\label{Z1Z4}
\left\{ \begin{array}{l}
M_\mu^2  - \left( {\frac{{\sqrt s }}{2} - k^0 } \right)^2  + \left( {k^1 } \right)^2  + \left( {k^2 } \right)^2  + \left( {P - k^3 } \right)^2  = 0, \\ 
M_G^2  - \left( {k^0 } \right)^2  + \left( { - P\sin \left( \theta  \right) - k^1 } \right)^2  + \left( {k^2 } \right)^2  + \left( {P\left( {1 - \cos \left( \theta  \right)} \right) - k^3 } \right)^2  = 0. \\ 
\end{array} \right.
\end{equation}
Тут якзазвичай $ \theta  $ позначено кут між векторами $ \vec P_1  $ і $ \vec P_3  $ . Або при нашому виборі вісів,  $ \theta$ є кут, який вектор $ \vec P_3  $  утворює з третьою координатною віссю (кут із першою віссю при обговореному виборі системи координат буде ${\pi  \mathord{\left/ 		{\vphantom {\pi  2}} \right. 		\kern-\nulldelimiterspace} 2} - \theta ,  $ а з другою -$ {\pi  \mathord{\left/		{\vphantom {\pi  2}} \right. 		\kern-\nulldelimiterspace} 2} $ )
\begin{equation}\label{Z1Z4preobr}
\left\{ \begin{array}{l}
\sqrt s k^0  - 2Pk_3  - \left( {k^0 } \right)^2  + \left( {k^1 } \right)^2  + \left( {k^2 } \right)^2  + \left( {k^3 } \right)^2  = 0, \\ 
M_G^2  + 2P^2 \left( {1 - \cos \left( \theta  \right)} \right) + 2P\sin \left( \theta  \right)k^1  + 2P\cos \left( \theta  \right)k^3  -  \\ 
- 2Pk_3  - \left( {k^0 } \right)^2  + \left( {k^1 } \right)^2  + \left( {k^2 } \right)^2  + \left( {k^3 } \right)^2  = 0. \\ 
\end{array} \right.
\end{equation}
Знов прирівнюючи вирази в лівих частинах рівностей \eqref{Z1Z4preobr}, як це було зроблено для \eqref{Z2Z3}, матимемо:
\begin{equation}\label{Z1Z4k0}
k^0  = \frac{{M_G^2  + 2P^2 \left( {1 - \cos \left( \theta  \right)} \right) + 2P\left( {\sin \left( \theta  \right)k^1  + \cos \left( \theta  \right)k^3 } \right)}}{{\sqrt s }}.
\end{equation}
Підставляючи це співвідношення в друге з рівнянь системи \eqref{Z1Z4} отримуємо
\begin{equation}\label{druge_rivnanna_Z1Z4}
\begin{split}
M_G^2  + 2P^2 \left( {1 - \cos \left( \theta  \right)} \right) + 2P\sin \left( \theta  \right)k^1  + 2P\cos \left( \theta  \right)k^3  - 2Pk^3  -  \\ 
- \left( {\frac{{M_G^2  + 2P^2 \left( {1 - \cos \left( \theta  \right)} \right) + 2P\sin \left( \theta  \right)k^1  + 2P\cos \left( \theta  \right)k^3 }}{{\sqrt s }}} \right)^2  +  \\ 
+ \left( {k^1 } \right)^2  + \left( {k^2 } \right)^2  + \left( {k^3 } \right)^2  = 0. \\ 
\end{split}
\end{equation}
Далі зручно перейти до нових координат
\begin{equation}\label{Z1Z4novicoordinsti}
\left\{ \begin{array}{l}
\sin \left( \theta  \right)k^1  + \cos \left( \theta  \right)k^3  = q^1 , \\ 
\cos \left( \theta  \right)k^1  - \sin \left( \theta  \right)k^3  = q^3 , \\ 
k^2  = q^2 . \\ 
\end{array} \right.
\end{equation}
В цих координатах рівняння \eqref{druge_rivnanna_Z1Z4} приймає вид
\begin{equation}\label{druge_rivnanna_Z1Z4_vq}
\begin{split}
M_G^2  + 2P^2 \left( {1 - \cos \left( \theta  \right)} \right) + 2Pq^1  - 2P\left( {\cos \left( \theta  \right)q^1  - \sin \left( \theta  \right)q^3 } \right) -  \\ 
- \left( {\frac{{M_G^2  + 2P^2 \left( {1 - \cos \left( \theta  \right)} \right) + 2Pq^1 }}{{\sqrt s }}} \right)^2  + \left( {q^1 } \right)^2  + \left( {q^2 } \right)^2  + \left( {q^3 } \right)^2  = 0. \\ 
\end{split}
\end{equation}

Виділяючи в цьому співвідношенні повні квадрати по $ q^1 $ i   $ q^3 $ після перетворень, отримаємо співвідношення:
\begin{equation}\label{Z1Z4zaperechenna}
\begin{split}
&M_G^2 \left( {1 - \frac{{M_G^2 }}{{4M_\mu^2 }}} \right) +  \\ 
&+ \frac{{4M_\mu^2 }}{s}\left( {q^1  + \frac{{P\left( {2M_\mu^2 \left( {1 - \cos \left( \theta  \right)} \right) - M_G^2 } \right)}}{{2M_\mu^2 }}} \right)^2  + \left( {q^2 } \right)^2  + \left( {q^3  + P\sin \left( \theta  \right)} \right)^2  = 0. \\ 
\end{split}
\end{equation}
З цього спввідношення видно, що й $ Z_1 $ і  $ Z_4 $ не можуть дорівнювати нулю одночасно.

Для другого і четвертого знаменників отримаємо систему
\begin{equation}\label{Z2Z4}
\begin{split}
\left\{ \begin{array}{l}
M_\mu^2  - \left( {{{\sqrt s } \mathord{\left/
			{\vphantom {{\sqrt s } 2}} \right.
			\kern-\nulldelimiterspace} 2} + k^0 } \right)^2  + \left( {k^1 } \right)^2  + \left( {k^2 } \right)^2  + \left( { - P + k^3 } \right)^2  = 0, \\ 
M_G^2  - \left( {k^0 } \right)^2  + \left( { - P\sin \left( \theta  \right) - k^1 } \right)^2  + \left( {k^2 } \right)^2  + \left( {P\left( {1 - \cos \left( \theta  \right)} \right) - k^3 } \right)^2  = 0. \\ 
\end{array} \right.
\end{split}
\end{equation}
яка знов відрізняється від \eqref{Z1Z4} лише заміною $k^0 $ на  $-k^0 .$ Тому вводячи замість  $k^0 $ нову змінну $q^0=-k^0 $ отримаємо систему, яка не має розв'язків. Це означає, що й система \eqref{Z2Z4} також розв'язків не має. 

Отже можемо зробити висновок, що жоден із знаменників, що відповідають вертикальним лініям не дорівнює одночасно нулю разом з жодним із знаменників, що відповідають горизонтальним лініям. Звідси можемо зробити висновок, що ані всі чотири знаменники, ані жодні три з них не можуть дорівнювати нулю одночасно. 

Розглянемо тепер можливість одночасного обертання в нуль для двох знаменників, що відповідають горизонтальним лініям діаграми рис.\ref{fig:naiprostihadiagrama1}, тобто першого і другого знаменників. Відповідна система рівнянь в системі центру мас вихідних частинок має вид:
\begin{equation}\label{Z1Z2}
\begin{split}
\left\{ \begin{array}{l}
 M_{\mu }^{2}-\left( {{\left( {\sqrt{s}}/{2}\;-{{k}^{0}} \right)}^{2}}-{{\left( {{k}^{1}} \right)}^{2}}-{{\left( {{k}^{2}} \right)}^{2}}-{{\left( P-{{k}^{3}} \right)}^{2}} \right)=0, \\ 
M_{\mu }^{2}-\left( {{\left( {\sqrt{s}}/{2}\;+{{k}^{0}} \right)}^{2}}-{{\left( {{k}^{1}} \right)}^{2}}-{{\left( {{k}^{2}} \right)}^{2}}-{{\left( -P+{{k}^{3}} \right)}^{2}} \right)=0. \\ 
\end{array} \right.
\end{split}
\end{equation}
 
 Враховуючи, що ${P^2} = {s \mathord{\left/ 		{\vphantom {s 4}} \right.  		\kern-\nulldelimiterspace} 4} - M_\mu ^2$
 , ці два рівняння можна переписати в виді:
 \begin{equation}\label{Z1Z2peretvorenna}
\begin{cases}
  \sqrt{s}{{k}^{0}}-{{\left( {{k}^{0}} \right)}^{2}}+{{\left( {{k}^{1}} \right)}^{2}}+{{\left( {{k}^{2}} \right)}^{2}}-2P{{k}^{3}}+{{\left( {{k}^{3}} \right)}^{2}}=0, \\ 
-\sqrt{s}{{k}^{0}}-{{\left( {{k}^{0}} \right)}^{2}}+{{\left( {{k}^{1}} \right)}^{2}}+{{\left( {{k}^{2}} \right)}^{2}}-2P{{k}^{3}}+{{\left( {{k}^{3}} \right)}^{2}}=0. \\ 
\end{cases}
\end{equation}

Віднімаючи від першого з цих рівнянь друге отримаємо $ 2\sqrt{s}{{k}^{0}}=0 $, звідки $k^0=0$. Підставляючи  $k^0=0$ в будь-яке з рівнянь системи \eqref{Z1Z2peretvorenna} після перетворень отримаємо рівняння $   {{\left( {{k}^{1}} \right)}^{2}}+{{\left( {{k}^{2}} \right)}^{2}}+{{\left( {{k}^{3}}-P \right)}^{2}}={{P}^{2}}. $

Отже серед множини стовпців виду 
\[ k=\left( \begin{matrix}
{{k}^{0}}  \\
{{k}^{1}}  \\
{{k}^{2}}  \\
{{k}^{3}}  \\
\end{matrix} \right) \]
можемо виділити підмножину тих стовпців, компоненти яких задовольняють системі рівнянь 
\begin{equation}\label{Pidmnojina1}
 \begin{cases}
{{k}^{0}}=0, \\ 
{{\left( {{k}^{1}} \right)}^{2}}+{{\left( {{k}^{2}} \right)}^{2}}+{{\left( {{k}^{3}}-P \right)}^{2}}={{P}^{2}}. \\ 
\end{cases} 
\end{equation}
Цю підмножину будемо далі називати <<підмножиною {1} >>. На цій підмножині перші два знаменникі дорівнюють нулю і тому ця підмножина дає внесок в інтеграл максимально можливого порядку ${1}/{{{\varepsilon }^{2}}}\;$. В наступному розділі ми розглянемо внесок підмножини {1} в інтеграл \eqref{fig:naiprostihadiagrama1}. 

Розглянемо тепер можливість рівності нулю третього і четвертого знаменників одночасно. Перед тим як розглядати відповідну систему рівнянь зручно зробити заміну змінних інтегрування. Після заміни маємо
\begin{equation}\label{Pisla_zamini}
\begin{split}
& {{A}_{1}}={G^{4}}\int{{{d}^{4}}q}\frac{1}{M_{\mu }^{2}-{{\left( \frac{1}{2}\left( {{P}_{1}}+{{P}_{3}} \right)-q \right)}^{2}}-i\varepsilon }\frac{1}{M_{\mu }^{2}-{{\left( \frac{1}{2}\left( {{P}_{4}}+{{P}_{2}} \right)+q \right)}^{2}}-i\varepsilon } \\ 
& \frac{1}{M_{G}^{2}-{{\left( \frac{1}{2}\left( {{P}_{1}}-{{P}_{3}} \right)+q \right)}^{2}}-i\varepsilon }\frac{1}{M_{G}^{2}-{{\left( \frac{1}{2}\left( {{P}_{1}}-{{P}_{3}} \right)-q \right)}^{2}}-i\varepsilon }. \\ 
\end{split}
\end{equation}

Також спростити подальший розгляд можна спеціальним вибором напрямків координатних вісів в системі центру мас вихідних частинок. Але можна ввести нові зручніші координати. Оскільки  \[\left| {{{\vec{P}}}_{1}} \right|=\left| {{{\vec{P}}}_{2}} \right|=\left| {{{\vec{P}}}_{3}} \right|=\left| {{{\vec{P}}}_{4}} \right|=\sqrt{{s}/{4}\;-M_{\mu }^{2}}\equiv P\] то паралелограм, побудований на векторах ${{\vec{P}}_{1}}$ і ${{\vec{P}}_{3}}$ є ромбом і його діагоналі, уздовж яких напрямлені вектори ${{\vec{P}}_{1}}+{{\vec{P}}_{3}}$ і ${{\vec{P}}_{1}}-{{\vec{P}}_{3}}$ є перпендикулярними. Тому зручно вісь 3 спрямувати уздовж вектору ${{\vec{P}}_{1}}-{{\vec{P}}_{3}}$, а вісь 1  – уздовж ${{\vec{P}}_{1}}+{{\vec{P}}_{3}}$. Вісь 2 залишається спрямувати перпендикулярно площині 1 і 3 так, щоб утворилася права система. Окрім того оскільки ${{\vec{P}}_{4}}=-{{\vec{P}}_{3}},{{\vec{P}}_{2}}=-{{\vec{P}}_{1}}$, вектор ${{\vec{P}}_{4}}+{{\vec{P}}_{2}}$, напрямлений протилежно ${{\vec{P}}_{1}}+{{\vec{P}}_{3}}$, тобто у від’ємному напрямку вісі 1. Тоді
\begin{equation}\label{P1P2P3P4}
\begin{split}
  & \frac{1}{2}\left( {{P}_{1}}-{{P}_{3}} \right)=\left( \begin{matrix}
0  \\
0  \\
0  \\
\frac{1}{2}\left| {{{\vec{P}}}_{1}}-{{{\vec{P}}}_{3}} \right|  \\
\end{matrix} \right),\frac{1}{2}\left( {{P}_{1}}+{{P}_{3}} \right)=\left( \begin{matrix}
{\sqrt{s}}/{2}\;  \\
\frac{1}{2}\left| {{{\vec{P}}}_{1}}+{{{\vec{P}}}_{3}} \right|  \\
0  \\
0  \\
\end{matrix} \right), \\ 
& \frac{1}{2}\left( {{P}_{2}}+{{P}_{4}} \right)=\left( \begin{matrix}
{\sqrt{s}}/{2}\;  \\
-\frac{1}{2}\left| {{{\vec{P}}}_{1}}+{{{\vec{P}}}_{3}} \right|  \\
0  \\
0  \\
\end{matrix} \right). \\ 
\end{split}
\end{equation}

Компоненти цих стовпців можна виразити через інваріанти Манделстама:

\begin{equation}\label{P1P2P3P4Mandelstam}
\begin{split}
  & \frac{1}{2}\left( {{P}_{1}}-{{P}_{3}} \right)=\left( \begin{matrix}
0  \\
0  \\
0  \\
{\sqrt{\left| t \right|}}/{2}\;  \\
\end{matrix} \right),\frac{1}{2}\left( {{P}_{1}}+{{P}_{3}} \right)=\left( \begin{matrix}
{\sqrt{s}}/{2}\;  \\
\sqrt{{{P}^{2}}-{\left| t \right|}/{4}\;}  \\
0  \\
0  \\
\end{matrix} \right), \\ 
& \frac{1}{2}\left( {{P}_{2}}+{{P}_{4}} \right)=\left( \begin{matrix}
{\sqrt{s}}/{2}\;  \\
-\sqrt{{{P}^{2}}-{\left| t \right|}/{4}\;}  \\
0  \\
0  \\
\end{matrix} \right). \\ 
\end{split}
\end{equation}

З урахуванням цих співвідношень, внесок діаграми рис.\ref{fig:naiprostihadiagrama1} може бути записаний в виді, в якому третій і четвертий знаменники приймають найбільш простий вид:
\begin{equation}\label{A1vnovixperem}
\begin{split}
   & {{A}_{1}}={{G}^{4}}\int\limits_{-\infty }^{+\infty }{d{{q}^{0}}}\int\limits_{-\infty }^{+\infty }{d{{q}^{1}}}\int\limits_{-\infty }^{+\infty }{d{{q}^{2}}}\int\limits_{-\infty }^{+\infty }{d{{q}^{3}}} \\ 
 & \frac{1}{M_{\mu }^{2}-{{\left( {\sqrt{s}}/{2}\;-{{q}^{0}} \right)}^{2}}+{{\left( \sqrt{{{P}^{2}}-{\left| t \right|}/{4}\;}-{{q}^{1}} \right)}^{2}}+{{\left( {{q}^{2}} \right)}^{2}}+{{\left( {{q}^{3}} \right)}^{2}}-i\varepsilon } \\ 
 & \frac{1}{M_{\mu }^{2}-{{\left( {\sqrt{s}}/{2}\;+{{q}^{0}} \right)}^{2}}+{{\left( \sqrt{{{P}^{2}}-{\left| t \right|}/{4}\;}-{{q}^{1}} \right)}^{2}}+{{\left( {{q}^{2}} \right)}^{2}}+{{\left( {{q}^{3}} \right)}^{2}}-i\varepsilon } \\ 
 & \frac{1}{M_{G}^{2}-{{\left( {{q}^{0}} \right)}^{2}}+{{\left( {{q}^{1}} \right)}^{2}}+{{\left( {{q}^{2}} \right)}^{2}}+{{\left( {\sqrt{\left| t \right|}}/{2}\;+{{q}^{3}} \right)}^{2}}-i\varepsilon } \\ 
 & \frac{1}{M_{G}^{2}-{{\left( {{q}^{0}} \right)}^{2}}+{{\left( {{q}^{1}} \right)}^{2}}+{{\left( {{q}^{2}} \right)}^{2}}+{{\left( {\sqrt{\left| t \right|}}/{2}\;-{{q}^{3}} \right)}^{2}}-i\varepsilon } .\\ 
\end{split}
\end{equation}
Відповідно, вимагаючи щоб третій і четвертий знаменники дорівнювали нулю, отримаємо систему рівнянь
\begin{equation}\label{Z3Z4}
\begin{cases}
& M_{G}^{2}+{\left| t \right|}/{4}\;-{{\left( {{q}^{0}} \right)}^{2}}+{{\left( {{q}^{1}} \right)}^{2}}+{{\left( {{q}^{2}} \right)}^{2}}+{{\left( {{q}^{3}} \right)}^{2}}+\sqrt{\left| t \right|}{{q}^{3}}=0, \\ 
& M_{G}^{2}+{\left| t \right|}/{4}\;-{{\left( {{q}^{0}} \right)}^{2}}+{{\left( {{q}^{1}} \right)}^{2}}+{{\left( {{q}^{2}} \right)}^{2}}+{{\left( {{q}^{3}} \right)}^{2}}-\sqrt{\left| t \right|}{{q}^{3}}=0. \\ 
\end{cases}.
\end{equation} 
Розв'язуючи цю систему отримаємо умови на підмножину множини стовпців 
\[ q=\left( \begin{matrix}
{{q}^{0}}  \\
{{q}^{1}}  \\
{{q}^{2}}  \\
{{q}^{3}}  \\
\end{matrix} \right),
 \]
по компонентах яких здійснюється інтегрування в \eqref{A1vnovixperem}, на якій дорівнюють нулю одночасно третій і четвертий знаменники. Цю підмножину, яку далі будемо називати << підмножиною {2}>> утворюють стовпці, компоненти яких задовольняють системі рівнянь:
 \begin{equation}\label{Z3Z4rozvazoc}
 \begin{cases}
& {{q}^{3}}=0,\left( \sqrt{\left| t \right|}\ne 0 \right) \\ 
& M_{G}^{2}+{\left| t \right|}/{4}\;-{{\left( {{q}^{0}} \right)}^{2}}+{{\left( {{q}^{1}} \right)}^{2}}+{{\left( {{q}^{2}} \right)}^{2}}=0. \\
 \end{cases}.
 \end{equation}  
Умова $ \left( \sqrt{\left| t \right|}\ne 0 \right) $ означає що випадок $t =0,$ ми збираємось розглядати як границю виразу \eqref{A1vnovixperem} при наближенні $t\to -0.$ Більш докладно цю границю ми розглянемо далі. Підмножина {2}, яка визначається рівняннями \eqref{Z3Z4rozvazoc} також як і розглянута вище підмножина {1} вносить в інтеграл \eqref{A1vnovixperem}, або в \eqref{Pisla_zamini} внесок найбільшого можливого порядку ${1}/{{{\varepsilon }^{2}}}\;$. Розглянемо тепер послідовно внески підмножин {1} і {2} в інтеграл, що відповідає діаграмі рис.\ref{fig:naiprostihadiagrama1}.
 
 \section{Розрахунок внеску підмножини {1} в інтеграл для діаграми рис.\ref{fig:naiprostihadiagrama1}}
 	Для опису внеску підмножини {1} зручно розглядати інтеграл в змінних інтегрування $ k, $ тобто в виді інтеграла \eqref{fi3_amplituda}, в той час як для підмножини {2} буде зручніше виходити з представлення інтегралу в виді \eqref{Pisla_zamini}. Обираючи вісі координат таким чином щоб вісь 3 була спрямована уздовж вектору $ \vec{P_1}, $ вісь 1 була спрямована перпендикулярно вісі 3 в тій площині, в якій лежать всі чотири вектори  $ \vec{P_1}, \vec{P_2}, \vec{P_3}, \vec{P_4},$ а вісь 2 була перпендикулярна цій площині і утворювала з вісями 1 і 3 праву систему координат, для інтеграла 
\eqref{fi3_amplituda} отримаємо вираз:
\begin{equation}\label{viraz_dla_A1}
\begin{split}
&A_1  = G^4 \int\limits_{ - \infty }^{ + \infty } {dk_0 } \int\limits_{ - \infty }^{ + \infty } {dk_1 } \int\limits_{ - \infty }^{ + \infty } {dk_2 } \int\limits_{ - \infty }^{ + \infty } {dk_3 }  \\ 
&\frac{1}{{M_\mu ^2  - \left( {\left( {\frac{{\sqrt s }}{2} - k^0 } \right)^2  - \left( {k^1 } \right)^2  - \left( {k^2 } \right)^2  - \left( {P - k^3 } \right)^2 } \right) - i\varepsilon }} \\ 
&\frac{1}{{M_\mu ^2  - \left( {\left( {\frac{{\sqrt s }}{2} + k^0 } \right)^2  - \left( {k^1 } \right)^2  - \left( {k^2 } \right)^2  - \left( { - P + k^3 } \right)^2 } \right) - i\varepsilon }} \\ 
&\frac{1}{{M_G^2  - \left( {k^0 } \right)^2  + \left( {k^1 } \right)^2  + \left( {k^2 } \right)^2  + \left( {k^3 } \right)^2  - i\varepsilon }} \\ 
&\frac{1}{{M_G^2  - \left( {k^0 } \right)^2  + \left( { - P\sin \left( \theta  \right) - k^1 } \right)^2  + \left( {k^2 } \right)^2  + \left( {P\left( {1 - \cos \left( \theta  \right)} \right) - k^3 } \right)^2  - i\varepsilon }} \\ 
\end{split}
\end{equation}
Після перетворень з урахуванням співвідношення $P^2  = {s \mathord{\left/ 		{\vphantom {s 4}} \right. 		\kern-\nulldelimiterspace} 4} - M_\mu ^2 $ інтеграл можна переписати в виді:
\begin{equation}\label{Pisla_peretvoren_viraz_1}
\begin{split}
&A_1  = G^4 \int\limits_{ - \infty }^{ + \infty } {dk_0 } \int\limits_{ - \infty }^{ + \infty } {dk_1 } \int\limits_{ - \infty }^{ + \infty } {dk_2 } \int\limits_{ - \infty }^{ + \infty } {dk_3 }  \times  \\ 
&\times \frac{1}{{ + \sqrt s k^0  - \left( {k^0 } \right)^2  + \left( {k^1 } \right)^2  + \left( {k^2 } \right)^2  - 2Pk^3  + \left( {k^3 } \right)^2  - i\varepsilon }} \times  \\ 
&\times \frac{1}{{ - \sqrt s k^0  - \left( {k^0 } \right)^2  + \left( {k^1 } \right)^2  + \left( {k^2 } \right)^2  - 2Pk^3  + \left( {k^3 } \right)^2  - i\varepsilon }} \times  \\ 
&\times \frac{1}{{M_G^2  - \left( {k^0 } \right)^2  + \left( {k^1 } \right)^2  + \left( {k^2 } \right)^2  + \left( {k^3 } \right)^2  - i\varepsilon }} \times  \\ 
&\times \frac{1}{{M_G^2  - \left( {k^0 } \right)^2  + \left( { - P\sin \left( \theta  \right) - k^1 } \right)^2  + \left( {k^2 } \right)^2  + \left( {P\left( {1 - \cos \left( \theta  \right)} \right) - k^3 } \right)^2  - i\varepsilon }}. \\ 
\end{split}
\end{equation}
Далі зручно перейти в інтегралі до сферичних координат $q,\theta _1 ,\varphi, $ які введемо у звичайний спосіб:
\begin{equation}\label{sfericni_koordinati}
q_3  = q\cos \left( {\theta _1 } \right),q_2  = q\sin \left( {\theta _1 } \right)\sin \left( \varphi  \right),q_1  = q\sin \left( {\theta _1 } \right)\cos \left( \varphi  \right)
\end{equation}
Після заміни \eqref{sfericni_koordinati} інтеграл \eqref{Pisla_peretvoren_viraz_1} приймає вид:
\begin{equation}\label{A1vsferichih_koordinatax}
\begin{split}
&A_1  =G^4 \int\limits_{ - \infty }^{ + \infty } {dk_0 } \int\limits_0^{ + \infty } {q^2 dq} \int\limits_0^\pi  {\sin \left( {\theta _1 } \right)d\theta _1 } \int\limits_0^{2\pi } d \varphi  \times  \\ 
&\times \frac{1}{{ + \sqrt s k^0  - \left( {k^0 } \right)^2  + q^2  - P^2    - i\varepsilon }}\frac{1}{{ - \sqrt s k^0  - \left( {k^0 } \right)^2  + q^2  - P^2  - i\varepsilon }} \times  \\ 
&\times \frac{1}{{M_G^2  - \left( {k^0 } \right)^2  + q^2  + 2Pq\cos \left( {\theta _1 } \right) + P^2  - i\varepsilon }} \times  \\ 
&\times \frac{1}{{M_G^2  - \left( {k^0 } \right)^2  + P^2  + q^2  + 2P\sin \left( \theta  \right)q\sin \left( {\theta _1 } \right)\cos \left( \varphi  \right) + 2P\cos \left( \theta  \right)q\cos \left( {\theta _1 } \right) - i\varepsilon }} \\ 
\end{split}
\end{equation}
Сенс переходу до сферичних координат полягає в тому що підмножина {1} тепер описується дуже простими рівняннями:
\begin{equation}\label{Prosti_rivnanna_dla_ pidmnojini1}
\begin{cases}
k^0  = 0, \\ 
q = P. \\ 
\end{cases}
\end{equation}

Отже, кожна точка області інтегрування визначається чотирма координатами
\[
\left( {\begin{array}{*{20}c}
	{k^0 }  \\
	q  \\
	{\theta _1 }  \\
	\varphi   \\
	\end{array}} \right)
\]
Якщо взяти довільну точку, що належить підмножині {1}, то координати ${k^0 }$ і $ q $ визначаються системою \eqref{Prosti_rivnanna_dla_ pidmnojini1}. Координати $ \theta _1 $ і $ \varphi$ можуть бути довільними в звичайних інтервалах для сферичних змінних. Отже змінюючи $ \theta _1 $ і $ \varphi$ ми будемо зсуватися з обраної точки уздовж підмножини {1}, а змінюючи ${k^0 }$ і $ q $ ми будемо зміщуватися із відповідної поверхні в області інтегрування в \eqref{A1vsferichih_koordinatax}. Отже уздовж поверхні \eqref{Prosti_rivnanna_dla_ pidmnojini1} підінтегральний вираз має максимум величиною ${1}/{{{\varepsilon }^{2}}}\;$. При цьому <<ширина максиму>> визначається залежністю підінтегрального виразу від змінних ${k^0 }$ і $ q $. Як обговорювалося раніше у вступі, ця  <<ширина максиму>> є по $ \varepsilon  $ для кожної з цих змінних. Це буде далі підтверджено більш докладними міркуваннями. Але перед тим буде зручно відділити внесок підмножини {1} від внеску підмножини {2}, бо інтеграл \eqref{A1vsferichih_koordinatax} містить внески обох підмножин. 

\begin{figure}
	\centering
	\includegraphics[scale=0.9]{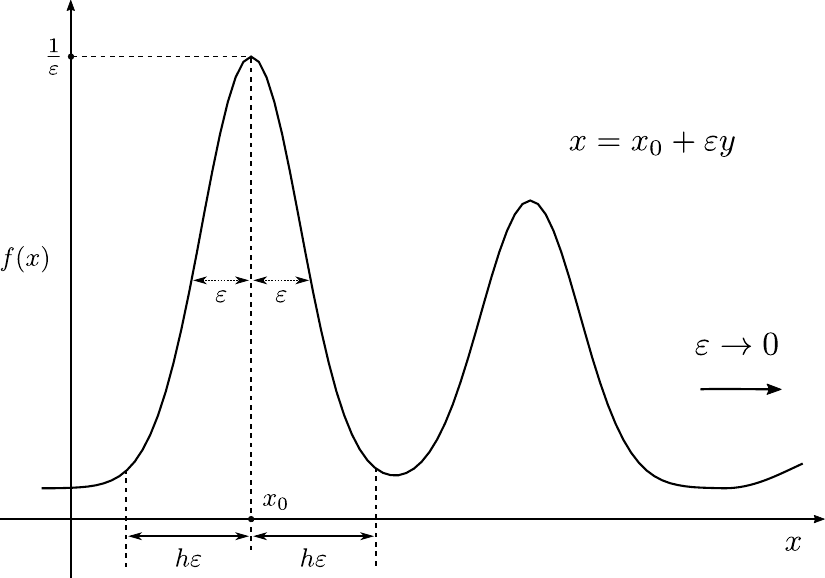}
	\caption[]{Схематична ілюстрація вихідної ситуації для розділення внесків підмножин {1} і {2} }
	\label{fig:peretvorenna_maximimu1}
\end{figure}

\begin{figure}
	\centering
	\includegraphics[scale=0.9]{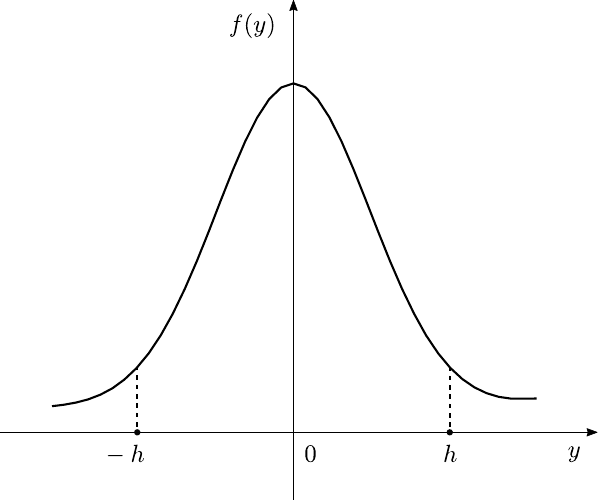}
	\caption[]{Внесок множини {1} після заміни змінних}
	\label{fig:peretvorenna_maximimu2}
\end{figure}

\begin{figure}
	\centering
	\includegraphics[scale=0.9]{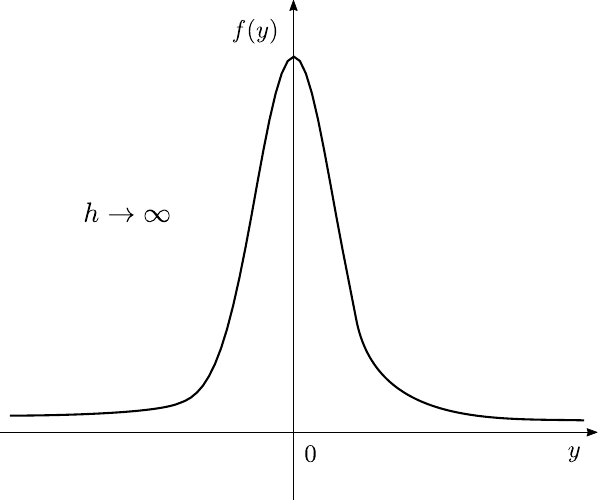}
	\caption[]{Внесок множини {1} після граничного переходу $h \to \infty $}
	\label{fig:peretvorenna_maximimu3}
\end{figure}

Для відділення внеску однієї підмножини від внеску іншої, розглянемо процедуру, яка проілюстрована на рис.\ref{fig:peretvorenna_maximimu1},рис.\ref{fig:peretvorenna_maximimu2},рис.\ref{fig:peretvorenna_maximimu3}.
	
Припустимо  $ \varepsilon  $ ще не наближено до нуля, але є малим. Припустимо, що ми маємо функцію як на рис.\ref{fig:peretvorenna_maximimu1} з двома максимумами. Відстань між максимумами не є нульовою (в нашому випадку це так внаслідок доведеного у попередньому розділі неперекриття поверхонь, що відповідають підмножинам {1} і {2}) . При цьому ця відстань не зменшується із зменшенням  $ \varepsilon  $ (неперекриття має місце за довільного значення $ \varepsilon  $ ). Тому якщо ми візьмемо за характерний масштаб $ \varepsilon$, то відстань між максимумами буде великою. 
Тоді можна обрати таке достатньо велике додатне число $ h ,$ що на кінцях проміжку $\left[ {{x}_{0}}-h\varepsilon ,{{x}_{0}}+h\varepsilon  \right]$ (рис.\ref{fig:peretvorenna_maximimu1}) функція встигає суттєво зменшитися у порівнянні з її значенням в точці максимуму. Тоді внеском цього максимуму в інтеграл від функції, будемо вважати інтеграл по проміжку $\left[ {{x}_{0}}-h\varepsilon ,{{x}_{0}}+h\varepsilon  \right].$ Тобто 
\begin{equation}\label{vneskidvohmaximumiv}
\int\limits_{-\infty }^{+\infty }{f\left( x \right)}dx\approx {{\left( \int\limits_{-\infty }^{+\infty }{f\left( x \right)}dx \right)}^{1\max }}+{{\left( \int\limits_{-\infty }^{+\infty }{f\left( x \right)}dx \right)}^{2\max }}
\end{equation}
Тут ${{\left( \int\limits_{-\infty }^{+\infty }{f\left( x \right)}dx \right)}^{1\max }}$i ${{\left( \int\limits_{-\infty }^{+\infty }{f\left( x \right)}dx \right)}^{2\max }}$ ми позначили внески відповідно першого і другого максимуму. Тоді, виходячи з сказаного:
\begin{equation}\label{vnesok1max}
{{\left( \int\limits_{-\infty }^{+\infty }{f\left( x \right)}dx \right)}^{1\max }}\approx \int\limits_{{{x}_{0}}-h\varepsilon }^{{{x}_{0}}+h\varepsilon }{f\left( x \right)}dx 
\end{equation}
Якщо тепер зробити заміну $x={{x}_{0}}+\varepsilon y,$ то міра інтегрування міститиме множник $ \varepsilon $ в чисельнику, в той час як максимальне значення підінтегральної функції міститиме аналогічний множник в знаменнику (рис.\ref{fig:peretvorenna_maximimu1}). Після того як ці множники скоротяться можемо перейти до границі $\varepsilon \to +0.$ Тоді як показано на (рис.\ref{fig:peretvorenna_maximimu1}) при переході до границі  $\varepsilon \to +0$ другий максимум зсувається на нескінченість, і в нас залишається інтеграл лише від першого максимуму в проміжку $\left[ -h,h \right]$ (рис.\ref{fig:peretvorenna_maximimu2})

Тепер ми можемо, знову ж таки в дусі методу Лапласа, наблизити $h$ до нескінченості (рис.\ref{fig:peretvorenna_maximimu3}), аналітично продовжуючи підінтегральну функцію на всю числову пряму, бо основний внесок в інтеграл буде вносити достатньо малий окіл точки максимуму. 
Граничний перехід $h\to +\infty $ позбавляє нас необхідності задавати конкретне значення $ h .$  

Застосуємо тепер описану процедуру до інтегралу \eqref{A1vsferichih_koordinatax}. Отже замість інтегралу \eqref{A1vsferichih_koordinatax}, розглянемо інтеграл по деякому околу підмножини {1}, що визначається формулами \eqref{Prosti_rivnanna_dla_ pidmnojini1}.
\begin{equation}\label{A1{1}}
\begin{split}
  & A_{1}^{\left\{ 1 \right\}}={{G}^{4}}\int\limits_{-h\varepsilon }^{h\varepsilon }{d{{k}_{0}}}\int\limits_{P-l\varepsilon }^{P+l\varepsilon }{{{q}^{2}}dq}\int\limits_{0}^{\pi }{\sin \left( {{\theta }_{1}} \right)d{{\theta }_{1}}}\int\limits_{0}^{2\pi }{d}\phi \times  \\ 
& \times \frac{1}{+\sqrt{s}{{k}^{0}}-{{\left( {{k}^{0}} \right)}^{2}}+\left( q-P \right)\left( q+P \right)-i\varepsilon }\frac{1}{-\sqrt{s}{{k}^{0}}-{{\left( {{k}^{0}} \right)}^{2}}+\left( q-P \right)\left( q+P \right)-i\varepsilon }\times  \\ 
& \times \frac{1}{M_{G}^{2}-{{\left( {{k}^{0}} \right)}^{2}}+{{q}^{2}}+2Pq\cos \left( {{\theta }_{1}} \right)+{{P}^{2}}-i\varepsilon }\times  \\ 
& \times \frac{1}{M_{G}^{2}-{{\left( {{k}^{0}} \right)}^{2}}+{{P}^{2}}+{{q}^{2}}+2P\sin \left( \theta  \right)q\sin \left( {{\theta }_{1}} \right)\cos \left( \phi  \right)+2P\cos \left( \theta  \right)q\cos \left( {{\theta }_{1}} \right)-i\varepsilon }. \\ 
\end{split}
\end{equation}
Тут позначення $ A_{1}^{\left\{ 1 \right\}} $ виражає той факт, що ми від всього інтегралу перейшли до інтегрування по деякому околу підмножини ${1}.$ Додатні числа $h$ і $l$ визначають ширину цього околу і далі будуть наближені до нескінченості у відповідності із описаною раніше процедурою. 

Перейдемо тепер від змінних ${{k}^{0}},q$ до нових змінних інтегрування $E,x$ за формулами
\begin{equation}\label{zamina_eps1}
{{k}_{0}}=\varepsilon E,q=P+\varepsilon x. 
\end{equation}

Тоді замість \eqref{A1{1}} отримаємо 
\begin{equation}\label{A1{1eps}}
\begin{split}
  & A_{1}^{\left\{ 1 \right\}}={{G}^{4}}\int\limits_{-h}^{h}{\varepsilon dE}\int\limits_{-l}^{l}{{{\left( P+\varepsilon x \right)}^{2}}\varepsilon dx}\int\limits_{0}^{\pi }{\sin \left( {{\theta }_{1}} \right)d{{\theta }_{1}}}\int\limits_{0}^{2\pi }{d}\phi \times  \\ 
& \times \frac{1}{\sqrt{s}\varepsilon E-{{\left( \varepsilon E \right)}^{2}}+\varepsilon x\left( P+\varepsilon x+P \right)-i\varepsilon }\times  \\ 
& \times \frac{1}{-\sqrt{s}\varepsilon E-{{\left( \varepsilon E \right)}^{2}}+\varepsilon x\left( P+\varepsilon x+P \right)-i\varepsilon }\times  \\ 
& \times \frac{1}{M_{G}^{2}-{{\left( \varepsilon E \right)}^{2}}+{{\left( P+\varepsilon x \right)}^{2}}+2P\left( P+\varepsilon x \right)\cos \left( {{\theta }_{1}} \right)+{{P}^{2}}-i\varepsilon }\times  \\ 
& \times \left( M_{G}^{2}-{{\left( \varepsilon E \right)}^{2}}+{{P}^{2}}+{{\left( P+\varepsilon x \right)}^{2}}+2P\sin \left( \theta  \right)\left( P+\varepsilon x \right)\sin \left( {{\theta }_{1}} \right)\cos \left( \phi  \right)+ \right. \\ 
& {{\left. +2P\left( P+\varepsilon x \right)\cos \left( \theta  \right)\cos \left( {{\theta }_{1}} \right)-i\varepsilon  \right)}^{-1}} \\ 
&  \\
\end{split}
\end{equation}
Перші два множника підінтегрального виразу містять в знаменниках по одному спільному множнику $\varepsilon $. Після їх винесення за дужки, вони можуть бути скорочені з двома множниками які виникли за рахунок перетворення  ${{dk}^{0}}$ і $dq.$ Після цього можемо переходити до границі $\varepsilon \to +0$ і отримаємо
\begin{equation}\label{ A_{1}eps0}
\begin{split}
& A_{1}^{\left\{ 1 \right\}}={{G}^{4}}{{P}^{2}}\int\limits_{-h}^{h}{dE}\int\limits_{-l}^{l}{dx}\frac{1}{\left( +\sqrt{s}E+2Px-i \right)}\frac{1}{\left( -\sqrt{s}E+2Px-i \right)}\times  \\ 
& \times \int\limits_{0}^{\pi }{\sin \left( {{\theta }_{1}} \right)d{{\theta }_{1}}}\frac{1}{M_{G}^{2}+2{{P}^{2}}\left( 1+\cos \left( {{\theta }_{1}} \right) \right)}\times  \\ 
& \times \int\limits_{0}^{2\pi }{d}\phi \frac{1}{M_{G}^{2}+2{{P}^{2}}\left( 1+\sin \left( \theta  \right)\sin \left( {{\theta }_{1}} \right)\cos \left( \phi  \right)+\cos \left( \theta  \right)\cos \left( {{\theta }_{1}} \right) \right)}. \\ 
\end{split}
\end{equation}
Як бачимо, після переходу до границі $\varepsilon \to +0$  маємо суттєве спрощення підінтегрального виразу яке дозволяє відділити інтегрування по $E$ і по $x$ від інтегрувань по решті змінних. Цей інтеграл зручно переписати в виді:
\begin{equation}\label{ poEipox}
\begin{split}
 & \int\limits_{-\infty }^{+\infty }{dE}\int\limits_{-\infty }^{+\infty }{dx}\frac{1}{\left( +\sqrt{s}E+2Px-i \right)}\frac{1}{\left( -\sqrt{s}E+2Px-i \right)}= \\ 
& =-\frac{1}{s}\int\limits_{-\infty }^{+\infty }{dx}\int\limits_{-\infty }^{+\infty }{dE}\frac{1}{\left( E-\left( -\frac{2P}{\sqrt{s}}x+\frac{i}{\sqrt{s}} \right) \right)}\frac{1}{\left( E-\left( \frac{2}{\sqrt{s}}Px-\frac{i}{\sqrt{s}} \right) \right)}. \\
\end{split}
\end{equation}
Після цього інтеграл по $E$ можна доповнити інтегралом по півколу в верхній, або нижній півплощині комплексної площини $E$ і до отриманого таким чином інтегралу по замкненому контуру застосувати теорему про лишки. Це призводить до наступного результату:
\begin{equation}\label{ pislaIntegruvannapoE}
\begin{split}
  & \int\limits_{-\infty }^{+\infty }{dE}\int\limits_{-\infty }^{+\infty }{dx}\frac{1}{\left( +\sqrt{s}E+2Px-i \right)}\frac{1}{\left( -\sqrt{s}E+2Px-i \right)}= \\ 
& =-\frac{\pi i}{s}\int\limits_{-\infty }^{+\infty }{dx}\frac{1}{-\frac{2P}{\sqrt{s}}x+\frac{i}{\sqrt{s}}} \\ 
\end{split}
\end{equation}
Подальше інтегрування не викликає проблем і призводить до результату
\begin{equation}\label{resultat_poE_i_po_x}
\begin{split}
  & \int\limits_{-\infty }^{+\infty }{dE}\int\limits_{-\infty }^{+\infty }{dx}\frac{1}{\left( +\sqrt{s}E+2Px-i \right)}\frac{1}{\left( -\sqrt{s}E+2Px-i \right)}= \\ 
& =\frac{\pi i}{2P\sqrt{s}}\ln \left( \frac{+\infty -\frac{i}{2P}}{-\infty -\frac{i}{2P}} \right)=-\frac{{{\pi }^{2}}}{2P\sqrt{s}}. \\ 
\end{split}
\end{equation}
В результаті маємо наступний вираз для внеску підмножини {1} 
\begin{equation}\label{viraz_dla_vnescu{1}}
\begin{split}
& A_{1}^{\left\{ 1 \right\}}=-{{G}^{4}}\frac{{{\pi }^{2}}P}{2\sqrt{s}}\int\limits_{0}^{\pi }{d{{\theta }_{1}}}\frac{\sin \left( {{\theta }_{1}} \right)}{M_{G}^{2}+2{{P}^{2}}\left( 1+\cos \left( {{\theta }_{1}} \right) \right)}\times  \\ 
& \times \int\limits_{0}^{2\pi }{d}\phi \frac{1}{M_{G}^{2}+2{{P}^{2}}\left( 1+\sin \left( \theta  \right)\sin \left( {{\theta }_{1}} \right)\cos \left( \phi  \right)+\cos \left( \theta  \right)\cos \left( {{\theta }_{1}} \right) \right)}. \\
\end{split}
\end{equation}
Далі зручно ввести заміну 
\begin{equation}\label{zamina_z}
\cos \left( {{\theta }_{1}} \right)=z, 
\end{equation}
після якої отримаємо:
\begin{equation}\label{pisla_zamini_z}
\begin{split}
  & A_{1}^{1\max }=-{{G}^{4}}\frac{{{\pi }^{2}}P}{2\sqrt{s}}\int\limits_{-1}^{1}{dz}\frac{1}{M_{G}^{2}+2{{P}^{2}}\left( 1+z \right)}\times  \\ 
& \times \int\limits_{0}^{2\pi }{d}\phi \frac{1}{M_{G}^{2}+2{{P}^{2}}\left( 1+\sin \left( \theta  \right)\sqrt{1-{{z}^{2}}}\cos \left( \phi  \right)+\cos \left( \theta  \right)z \right)} .\\ 
\end{split}
\end{equation}
Інтеграл по $ \phi$ може бути перетворений наступним чином:
\begin{equation}\label{peretvorenna_integral_po_fi}
\begin{split}
& \int\limits_{0}^{2\pi }{d}\phi \frac{1}{M_{G}^{2}+2{{P}^{2}}\left( 1+\sin \left( \theta  \right)\sqrt{1-{{z}^{2}}}\cos \left( \phi  \right)+\cos \left( \theta  \right)z \right)}= \\ 
& =\int\limits_{-\pi }^{\pi }{d}\phi \frac{1}{M_{G}^{2}+2{{P}^{2}}\left( 1+\sin \left( \theta  \right)\sqrt{1-{{z}^{2}}}\cos \left( \phi  \right)+\cos \left( \theta  \right)z \right)}. \\  
\end{split}
\end{equation}

Далі, після заміни змінної $ \operatorname{tg}\left( {\phi }/{2}\; \right)=x $ 
\begin{equation}\label{integral_po_fi}
\begin{split}
  & \int\limits_{-\pi }^{\pi }{d}\phi \frac{1}{M_{G}^{2}+2{{P}^{2}}\left( 1+\sin \left( \theta  \right)\sqrt{1-{{z}^{2}}}\cos \left( \phi  \right)+\cos \left( \theta  \right)z \right)}= \\ 
& =\frac{-8\pi }{\left( M_{G}^{2}+2{{P}^{2}}\left( 1+\cos \left( \theta  \right)z-\sin \left( \theta  \right)\sqrt{1-{{z}^{2}}} \right) \right)}\times  \\ 
& \times \sqrt{\frac{M_{G}^{2}+2{{P}^{2}}\left( 1+\cos \left( \theta  \right)z+\sin \left( \theta  \right)\sqrt{1-{{z}^{2}}} \right)}{\left( M_{G}^{2}+2{{P}^{2}}\left( 1+\cos \left( \theta  \right)z-\sin \left( \theta  \right)\sqrt{1-{{z}^{2}}} \right) \right)}} \\ 
\end{split}
\end{equation}

В результаті розрахунок внеску підмножини 1 звівся до одновимірного інтегралу в скінчених границях 
\begin{equation}\label{rezultat_A1}
\begin{split}
  & A_{1}^{\left\{ 1 \right\}}=G^{4}\frac{4{{\pi }^{3}}P}{\sqrt{s}}\int\limits_{-1}^{1}{dz}\frac{1}{M_{G}^{2}+2{{P}^{2}}\left( 1+z \right)}\times  \\ 
& \times \frac{\sqrt{M_{G}^{2}+2{{P}^{2}}\left( 1+\cos \left( \theta  \right)z+\sin \left( \theta  \right)\sqrt{1-{{z}^{2}}} \right)}}{{{\left( M_{G}^{2}+2{{P}^{2}}\left( 1+\cos \left( \theta  \right)z-\sin \left( \theta  \right)\sqrt{1-{{z}^{2}}} \right) \right)}^{3/2}}}. \\
\end{split}
\end{equation}
Цей інетеграл вже може бути розраховано, наприклад, методом трапецій. 

\section{Розрахунок внеску підмножини {2} в інтеграл для діаграми рис.\ref{fig:naiprostihadiagrama1}}
Для врахування внеску підмножини 2, як вже зазначалося, зручно виходити з виразу \eqref{A1vnovixperem}. Підінтегральний вираз в інтегралі \eqref{A1vnovixperem} є парною функцією від ${{q}^{0}}$. Тому:
\begin{equation}\label{A1parn }
\begin{split}
  & {{A}_{1}}=2G^4\int\limits_{0}^{+\infty }{d{{q}^{0}}}\int\limits_{-\infty }^{+\infty }{d{{q}^{1}}}\int\limits_{-\infty }^{+\infty }{d{{q}^{2}}}\int\limits_{-\infty }^{+\infty }{d{{q}^{3}}}\times  \\ 
& \times \frac{1}{\sqrt{s}{{q}^{0}}-{{\left( {{q}^{0}} \right)}^{2}}+{{\left( {{q}^{1}} \right)}^{2}}+{{\left( {{q}^{2}} \right)}^{2}}+{{\left( {{q}^{3}} \right)}^{2}}-{\left| t \right|}/{4}\;-2{{q}^{1}}\sqrt{{{P}^{2}}-{\left| t \right|}/{4}\;}-i\varepsilon }\times  \\ 
& \times \frac{1}{-\sqrt{s}{{q}^{0}}-{{\left( {{q}^{0}} \right)}^{2}}+{{\left( {{q}^{1}} \right)}^{2}}+{{\left( {{q}^{2}} \right)}^{2}}+{{\left( {{q}^{3}} \right)}^{2}}-{\left| t \right|}/{4}\;-2{{q}^{1}}\sqrt{{{P}^{2}}-{\left| t \right|}/{4}\;}-i\varepsilon }\times  \\ 
& \times \frac{1}{M_{G}^{2}-{{\left( {{q}^{0}} \right)}^{2}}+{{\left( {{q}^{1}} \right)}^{2}}+{{\left( {{q}^{2}} \right)}^{2}}+{{\left( {{q}^{3}} \right)}^{2}}+{\left| t \right|}/{4}\;+\sqrt{\left| t \right|}{{q}^{3}}-i\varepsilon }\times  \\ 
& \times \frac{1}{M_{G}^{2}-{{\left( {{q}^{0}} \right)}^{2}}+{{\left( {{q}^{1}} \right)}^{2}}+{{\left( {{q}^{2}} \right)}^{2}}+{{\left( {{q}^{3}} \right)}^{2}}+{\left| t \right|}/{4}\;-\sqrt{\left| t \right|}{{q}^{3}}-i\varepsilon }. \\ 
\end{split}
\end{equation}
Знов, як і в попередньому розділі введемо сферичні координати \eqref{sfericni_koordinati} з наступною заміною \eqref{zamina_z}. Після таких перетворень, матимемо:
\begin{equation}\label{A1parn_sfer_perem }
\begin{split}
& {{A}_{1}}=2{{G}^{4}}\int\limits_{0}^{+\infty }{d{{q}^{0}}}\int\limits_{0}^{+\infty }{{{q}^{2}}dq}\int\limits_{-1}^{1}{dz}\int\limits_{0}^{2\pi }{d\phi }\times  \\ 
& \times \frac{1}{\sqrt{s}{{q}^{0}}-{{\left( {{q}^{0}} \right)}^{2}}+{{\left( q \right)}^{2}}-{\left| t \right|}/{4}\;-2q\sqrt{1-{{z}^{2}}}\cos \left( \phi  \right)\sqrt{{{P}^{2}}-{\left| t \right|}/{4}\;}-i\varepsilon }\times  \\ 
& \times \frac{1}{-\sqrt{s}{{q}^{0}}-{{\left( {{q}^{0}} \right)}^{2}}+{{\left( q \right)}^{2}}-{\left| t \right|}/{4}\;-2q\sqrt{1-{{z}^{2}}}\cos \left( \phi  \right)\sqrt{{{P}^{2}}-{\left| t \right|}/{4}\;}-i\varepsilon }\times  \\ 
& \times \frac{1}{M_{G}^{2}+{\left| t \right|}/{4}\;+{{\left( q \right)}^{2}}-{{\left( {{q}^{0}} \right)}^{2}}+\sqrt{\left| t \right|}qz-i\varepsilon }\times  \\ 
& \times \frac{1}{M_{G}^{2}+{\left| t \right|}/{4}\;+{{\left( q \right)}^{2}}-{{\left( {{q}^{0}} \right)}^{2}}-\sqrt{\left| t \right|}qz-i\varepsilon }. \\ 
\end{split}
\end{equation}
Підмножина 2 у введених змінних задається системою рівнянь
\begin{equation}\label{Pidmnojina2sfer_perem}
\begin{cases}
& z=0\left( \sqrt{\left| t \right|}\ne 0 \right), \\ 
& {{q}^{0}}=\sqrt{M_{G}^{2}+\frac{\left| t \right|}{4}+{{\left( q \right)}^{2}}}. \\
\end{cases}
\end{equation}
Внесок околу підмножини 2 в інтеграл \eqref{A1parn_sfer_perem }(позначатимемо його далі $ A_{1}^{\left\{ 2 \right\}} $) можемо записати в виді:
\begin{equation}\label{A1{2}}
\begin{split}
A_{1}^{\left\{ 2 \right\}}=2{{G}^{4}}\int\limits_{\sqrt{M_{G}^{2}+\frac{\left| t \right|}{4}+{{\left( q \right)}^{2}}}-h\varepsilon }^{\sqrt{M_{G}^{2}+\frac{\left| t \right|}{4}+{{\left( q \right)}^{2}}}+h\varepsilon }{d{{q}^{0}}}\int\limits_{0}^{+\infty }{{{q}^{2}}dq}\int\limits_{-l\varepsilon }^{l\varepsilon }{dz}\int\limits_{0}^{2\pi }{d\phi }f\left( {{q}^{0}},q,z,\phi  \right)g\left( {{q}^{0}},q,z,\phi  \right).
\end{split}
\end{equation}
Тут введено позначення
\begin{equation}\label{Poznachenna_fig}
\begin{split}
& f\left( {{q}^{0}},q,z,\phi  \right)\equiv \frac{1}{\sqrt{s}{{q}^{0}}-{{\left( {{q}^{0}} \right)}^{2}}+{{\left( q \right)}^{2}}-{\left| t \right|}/{4}\;-2q\sqrt{1-{{z}^{2}}}\cos \left( \phi  \right)\sqrt{{{P}^{2}}-{\left| t \right|}/{4}\;}-i\varepsilon }\times  \\ 
& \times \frac{1}{-\sqrt{s}{{q}^{0}}-{{\left( {{q}^{0}} \right)}^{2}}+{{\left( q \right)}^{2}}-{\left| t \right|}/{4}\;-2q\sqrt{1-{{z}^{2}}}\cos \left( \phi  \right)\sqrt{{{P}^{2}}-{\left| t \right|}/{4}\;}-i\varepsilon }, \\ 
& g\left( {{q}^{0}},q,z,\phi  \right)=\frac{1}{M_{G}^{2}+{\left| t \right|}/{4}\;+{{\left( q \right)}^{2}}-{{\left( {{q}^{0}} \right)}^{2}}+\sqrt{\left| t \right|}qz-i\varepsilon }\times  \\ 
& \times \frac{1}{M_{G}^{2}+{\left| t \right|}/{4}\;+{{\left( q \right)}^{2}}-{{\left( {{q}^{0}} \right)}^{2}}-\sqrt{\left| t \right|}qz-i\varepsilon }. \\
\end{split}
\end{equation}
Інтеграл \eqref{A1{2}} перепишемо в виді:
\begin{equation}\label{PreobrazovanieA1{2}}
\begin{split}
& A_{1}^{\left\{ 2 \right\}}=2{{G}^{4}}\int\limits_{\sqrt{M_{G}^{2}+\frac{\left| t \right|}{4}+{{\left( q \right)}^{2}}}-h\varepsilon }^{\sqrt{M_{G}^{2}+\frac{\left| t \right|}{4}+{{\left( q \right)}^{2}}}+h\varepsilon }{d{{q}^{0}}}\int\limits_{0}^{+\infty }{{{q}^{2}}dq}\int\limits_{-l\varepsilon }^{l\varepsilon }{dz}\int\limits_{0}^{2\pi }{d\phi }\times  \\ 
& \times \left( f\left( {{q}^{0}},q,z,\phi  \right)-f\left( {{q}^{0}}=\sqrt{M_{G}^{2}+\frac{\left| t \right|}{4}+{{\left( q \right)}^{2}}},q,z=0,\phi  \right) \right)g\left( {{q}^{0}},q,z,\phi  \right)+ \\ 
& +2{{G}^{4}}\int\limits_{0}^{+\infty }{{{q}^{2}}dq}\int\limits_{0}^{2\pi }{d\phi }\int\limits_{\sqrt{M_{G}^{2}+\frac{\left| t \right|}{4}+{{\left( q \right)}^{2}}}-h\varepsilon }^{\sqrt{M_{G}^{2}+\frac{\left| t \right|}{4}+{{\left( q \right)}^{2}}}+h\varepsilon }{d{{q}^{0}}}\int\limits_{-l\varepsilon }^{l\varepsilon }{dz}\times  \\ 
& \times f\left( {{q}^{0}}=\sqrt{M_{G}^{2}+\frac{\left| t \right|}{4}+{{\left( q \right)}^{2}}},q,z=0,\phi  \right)g\left( {{q}^{0}},q,z,\phi  \right). \\ 
\end{split}
\end{equation}
Тобто ми додали і відняли від множника $f$ його значення на підмножині 2. Далі, зробимо заміну
\begin{equation}\label{Zamina_na_pidmnojini2}
{{q}^{0}}=\sqrt{M_{G}^{2}+\frac{\left| t \right|}{4}+{{\left( q \right)}^{2}}}+\varepsilon E,z=\varepsilon x.
\end{equation}
Після заміни \eqref{Zamina_na_pidmnojini2}, замість \eqref{PreobrazovanieA1{2}} отримаємо
\begin{equation}\label{pisla_zamini_A1{2} }
\begin{split}
& A_{1}^{\left\{ 2 \right\}}=2{{G}^{4}}\int\limits_{0}^{+\infty }{{{q}^{2}}dq}\int\limits_{0}^{2\pi }{d\phi }\int\limits_{-h}^{h}{\varepsilon dE}\int\limits_{-l}^{l}{\varepsilon dx}\times  \\ 
& \times \left( f\left( {{q}^{0}}=\sqrt{M_{G}^{2}+\frac{\left| t \right|}{4}+{{\left( q \right)}^{2}}}+\varepsilon E,q,z=\varepsilon x,\phi  \right)-f\left( {{q}^{0}}=\sqrt{M_{G}^{2}+\frac{\left| t \right|}{4}+{{\left( q \right)}^{2}}},q,z=0,\phi  \right) \right)\times  \\ 
& \times g\left( {{q}^{0}}=\sqrt{M_{G}^{2}+\frac{\left| t \right|}{4}+{{\left( q \right)}^{2}}}+\varepsilon E,q,z=\varepsilon x,\phi  \right)+ \\ 
& +2{{G}^{4}}\int\limits_{0}^{+\infty }{{{q}^{2}}dq}\int\limits_{0}^{2\pi }{d\phi }\int\limits_{-h}^{h}{\varepsilon dE}\int\limits_{-l}^{l}{\varepsilon dx}\times  \\ 
& \times f\left( {{q}^{0}}=\sqrt{M_{G}^{2}+\frac{\left| t \right|}{4}+{{\left( q \right)}^{2}}},q,z=0,\phi  \right)g\left( {{q}^{0}}=\sqrt{M_{G}^{2}+\frac{\left| t \right|}{4}+{{\left( q \right)}^{2}}}+\varepsilon E,q,z=\varepsilon x,\phi  \right). \\ 
\end{split}
\end{equation}

Врахуємо, що на підмножині 2 
\begin{equation}\label{g0}
\begin{split}
& g\left( {{q}^{0}}=\sqrt{M_{G}^{2}+\frac{\left| t \right|}{4}+{{\left( q \right)}^{2}}}+\varepsilon E,q,z=\varepsilon x,\phi  \right)= \\ 
& =\frac{1}{{{\varepsilon }^{2}}}{{g}_{0}}\left( {{q}^{0}}=\sqrt{M_{G}^{2}+\frac{\left| t \right|}{4}+{{\left( q \right)}^{2}}}+\varepsilon E,q,z=\varepsilon x,\phi  \right) .\\
\end{split}
\end{equation}
При цьому, функція $g0$ має скінчену границю при $\varepsilon \to +0$.

Розкладаючи в ряд Тейлора різницю у першому доданку формули \eqref{pisla_zamini_A1{2} } матимемо:
\begin{equation}\label{Tejlor}
\begin{split}
  & A_{1}^{\left\{ 2 \right\}}=2{{G}^{4}}\int\limits_{0}^{+\infty }{{{q}^{2}}dq}\int\limits_{0}^{2\pi }{d\phi }\int\limits_{-h}^{h}{dE}\int\limits_{-l}^{l}{dx}\times  \\ 
& \times \left( \frac{\partial f}{\partial {{q}^{0}}}\left( {{q}^{0}}=\sqrt{M_{G}^{2}+\frac{\left| t \right|}{4}+{{\left( q \right)}^{2}}},q,z=0,\phi  \right)\varepsilon E+ \right. \\ 
& \left. +\frac{\partial f}{\partial z}\left( {{q}^{0}}=\sqrt{M_{G}^{2}+\frac{\left| t \right|}{4}+{{\left( q \right)}^{2}}},q,z=0,\phi  \right)\varepsilon x+\left( {} \right){{\varepsilon }^{2}}+\left( {} \right){{\varepsilon }^{3}}+\ldots  \right)\times  \\ 
& \times {{g}_{0}}\left( {{q}^{0}}=\sqrt{M_{G}^{2}+\frac{\left| t \right|}{4}+{{\left( q \right)}^{2}}}+\varepsilon E,q,z=\varepsilon x,\phi  \right)+ \\ 
& +2{{G}^{4}}\int\limits_{0}^{+\infty }{{{q}^{2}}dq}\int\limits_{0}^{2\pi }{d\phi }\int\limits_{-h}^{h}{dE}\int\limits_{-l}^{l}{dx}\times  \\ 
& \times f\left( {{q}^{0}}=\sqrt{M_{G}^{2}+\frac{\left| t \right|}{4}+{{\left( q \right)}^{2}}},q,z=0,\phi  \right){{g}_{0}}\left( {{q}^{0}}=\sqrt{M_{G}^{2}+\frac{\left| t \right|}{4}+{{\left( q \right)}^{2}}}+\varepsilon E,q,z=\varepsilon x,\phi  \right) .\\
\end{split}
\end{equation} 
Граничний перехід $\varepsilon \to +0$ дозволяє суттєво спростити вираз \eqref{Tejlor}. Цей вираз приймає вид:
\begin{equation}\label{Pisla_epsilon_nul}
\begin{split}
& A_{1}^{\left\{ 2 \right\}}=-{{G}^{4}}2\int\limits_{0}^{+\infty }{{{q}^{2}}dq}\int\limits_{0}^{2\pi }{d\phi }\int\limits_{-h}^{h}{E}\int\limits_{-l}^{l}{dx}\times  \\ 
& \times \frac{1}{s\left( M_{G}^{2}+\frac{\left| t \right|}{4}+{{\left( q \right)}^{2}} \right)-{{\left( M_{G}^{2}+\frac{\left| t \right|}{2}+2q\cos \left( \phi  \right)\sqrt{{{P}^{2}}-\frac{\left| t \right|}{4}} \right)}^{2}}}\times  \\ 
& \times \frac{1}{2\sqrt{M_{G}^{2}+\frac{\left| t \right|}{4}+{{\left( q \right)}^{2}}}E-\sqrt{\left| t \right|}qx+i}\times  \\ 
& \times \frac{1}{2\sqrt{M_{G}^{2}+\frac{\left| t \right|}{4}+{{\left( q \right)}^{2}}}E+\sqrt{\left| t \right|}qx+i}. \\ 
\end{split}
\end{equation}
Виконуючи інтегрування по $E,$ отримаємо:
\begin{equation}\label{integruvanna_po_E}
\begin{split}
& A_{1}^{\{2\}}=2{{G}^{4}}\int\limits_{0}^{+\infty }{{{q}^{2}}dq}\int\limits_{0}^{2\pi }{d\phi }\int\limits_{-l}^{l}{dx}\frac{1}{s\left( M_{G}^{2}+\frac{\left| t \right|}{4}+{{\left( q \right)}^{2}} \right)-{{\left( M_{G}^{2}+\frac{\left| t \right|}{2}+2q\cos \left( \phi  \right)\sqrt{{{P}^{2}}-\frac{\left| t \right|}{4}} \right)}^{2}}}\times  \\ 
& \times \frac{1}{2\sqrt{\left| t \right|}qx2\sqrt{M_{G}^{2}+\frac{\left| t \right|}{4}+{{\left( q \right)}^{2}}}}\left( \ln \left( \frac{h+\frac{i-\sqrt{\left| t \right|}qx}{2\sqrt{M_{G}^{2}+\frac{\left| t \right|}{4}+{{\left( q \right)}^{2}}}}}{h+\frac{i+\sqrt{\left| t \right|}qx}{2\sqrt{M_{G}^{2}+\frac{\left| t \right|}{4}+{{\left( q \right)}^{2}}}}} \right)-\ln \left( \frac{-h+\frac{i-\sqrt{\left| t \right|}qx}{2\sqrt{M_{G}^{2}+\frac{\left| t \right|}{4}+{{\left( q \right)}^{2}}}}}{-h+\frac{i+\sqrt{\left| t \right|}qx}{2\sqrt{M_{G}^{2}+\frac{\left| t \right|}{4}+{{\left( q \right)}^{2}}}}} \right) \right). \\ 
\end{split}
\end{equation}
Далі ми збираємось $ h $ наблизити до нескінченості. Будемо перед цим вважати $ h $ великою додатною величиною. Тоді:
\begin{equation}\label{Razlijenie_po_h}
\begin{split}
& \ln \left( \frac{1+\frac{i-\sqrt{\left| t \right|}qx}{2h\sqrt{M_{G}^{2}+\frac{\left| t \right|}{4}+{{\left( q \right)}^{2}}}}}{1+\frac{i+\sqrt{\left| t \right|}qx}{2h\sqrt{M_{G}^{2}+\frac{\left| t \right|}{4}+{{\left( q \right)}^{2}}}}} \right)-\ln \left( \frac{1-\frac{i-\sqrt{\left| t \right|}qx}{2h\sqrt{M_{G}^{2}+\frac{\left| t \right|}{4}+{{\left( q \right)}^{2}}}}}{1-\frac{i+\sqrt{\left| t \right|}qx}{2h\sqrt{M_{G}^{2}+\frac{\left| t \right|}{4}+{{\left( q \right)}^{2}}}}} \right)= \\ 
& =-\frac{2\sqrt{\left| t \right|}qx}{h\sqrt{M_{G}^{2}+\frac{\left| t \right|}{4}+{{\left( q \right)}^{2}}}} \\
\end{split}
\end{equation}
З урахуванням цього результату легко виконати в \eqref{integruvanna_po_E} інтегрування по $ x $. Його результат має вид
\begin{equation}\label{A1{2}lh}
\begin{split}
& A_{1}^{\left\{ 2 \right\}}=-2{{G}^{4}}\frac{l}{h}\int\limits_{0}^{+\infty }{{{q}^{2}}dq}\frac{1}{\left( M_{G}^{2}+\frac{\left| t \right|}{4}+{{\left( q \right)}^{2}} \right)}\times  \\ 
& \times \int\limits_{0}^{2\pi }{d\phi }\frac{1}{s\left( M_{G}^{2}+\frac{\left| t \right|}{4}+{{\left( q \right)}^{2}} \right)-{{\left( M_{G}^{2}+\frac{\left| t \right|}{2}+2q\cos \left( \phi  \right)\sqrt{{{P}^{2}}-\frac{\left| t \right|}{4}} \right)}^{2}}}. \\ 
\end{split}
\end{equation}
Для того, щоб позбавитись залежності від $ l $ і $ h $ покладемо
\begin{equation}\label{Granica}
l=h\to +\infty.
\end{equation}

Для подальшого розрахунку зручно ввести позначення 
\begin{equation}\label{mkvadrat_ot_t}
M_{G}^{2}+\frac{\left| t \right|}{4}\equiv {{m}^{2}}\left( t \right),
\end{equation}
а також замість змінної інтегрування $ q $ ввести нову змінну інтегрування $y: $
\begin{equation}\label{yvmestoq}
q=m\left( t \right)\operatorname{sh}\left( y \right)
\end{equation}
Після цих перетворень отримаємо
\begin{equation}\label{A1{2}po_fi_po_y}
\begin{split}
  & A_{1}^{\left\{ 2 \right\}}=-2\frac{{{G}^{4}}}{m\left( t \right)}\int\limits_{0}^{2\pi }{d\phi }\times  \\ 
& \times \int\limits_{0}^{+\infty }{dy\frac{{{\operatorname{sh}}^{2}}\left( y \right)}{\operatorname{ch}\left( y \right)}}\frac{1}{\left( s{{\operatorname{ch}}^{2}}\left( y \right)-{{\left( m\left( t \right)+{\left| t \right|}/{4m\left( t \right)}\;+2\operatorname{sh}\left( y \right)\cos \left( \phi  \right)\sqrt{{{P}^{2}}-{\left| t \right|}/{4}\;} \right)}^{2}} \right)}. \\ 
\end{split}
\end{equation}

Щоб перейти до інтегрування в скінчених границях, яке було б зручно виконати чисельно зробимо заміну:
\begin{equation}\label{zamina_u}
\exp \left( -y \right)=u.
\end{equation}
Після цієї заміни отримаємо вираз:
\begin{equation}\label{integral_cherez_u}
\begin{split}
  & A_{1}^{\left\{ 2 \right\}}=-4\frac{{{G}^{4}}}{m\left( t \right)}\int\limits_{0}^{2\pi }{d\phi }\int\limits_{0}^{1}{du\frac{{{\left( 1-{{u}^{2}} \right)}^{2}}}{\left( 1+{{u}^{2}} \right)}}\times  \\ 
& \times \frac{1}{\left( \sqrt{s}\left( 1+{{u}^{2}} \right)-\left( 1-{{u}^{2}} \right)\cos \left( \phi  \right)\sqrt{s-\left| t \right|-4M_{\mu }^{2}}-2m\left( t \right)u\left( 1+{\left| t \right|}/{4{{m}^{2}}\left( t \right)}\; \right) \right)}\times  \\ 
& \times \frac{1}{\left( \sqrt{s}\left( 1+{{u}^{2}} \right)+\left( 1-{{u}^{2}} \right)\cos \left( \phi  \right)\sqrt{s-\left| t \right|-4M_{\mu }^{2}}+2m\left( t \right)u\left( 1+{\left| t \right|}/{4{{m}^{2}}\left( t \right)}\; \right) \right)}. \\ 
\end{split}
\end{equation}
Отже, інтеграл по $u$ розраховується в скінчених границях і підінтегральний вираз не має особливостей на проміжку інтегрування. Розглянемо тепер интегрування по $ \phi $. Для цього перепишемо інтеграл в виді:
\begin{equation}\label{integririvanie_po_fi_1}
\begin{split}
& A_{1}^{\left\{ 2 \right\}}=\frac{4{{G}^{4}}}{m\left( t \right)}\int\limits_{0}^{1}{du\frac{{{\left( 1-{{u}^{2}} \right)}^{2}}}{\left( 1+{{u}^{2}} \right)}\frac{1}{{{\left( \left( 1-{{u}^{2}} \right)\sqrt{s-\left| t \right|-4M_{\mu }^{2}} \right)}^{2}}}}\times  \\ 
& \times \int\limits_{0}^{2\pi }{d\phi }\frac{1}{\left( \cos \left( \phi  \right)-{{z}_{1}} \right)}\frac{1}{\left( \cos \left( \phi  \right)-{{z}_{2}} \right)}, \\ 
\end{split}
\end{equation}
де 
\begin{equation}\label{z1iz2}
\begin{split}
& {{z}_{1}}=\frac{\sqrt{s}\left( 1+{{u}^{2}} \right)-2m\left( t \right)u\left( 1+\frac{\left| t \right|}{4{{m}^{2}}\left( t \right)} \right)}{\left( 1-{{u}^{2}} \right)\sqrt{s-\left| t \right|-4M_{\mu }^{2}}}, \\ 
& {{z}_{2}}=-\frac{\sqrt{s}\left( 1+{{u}^{2}} \right)+2m\left( t \right)u\left( 1+\frac{\left| t \right|}{4{{m}^{2}}\left( t \right)} \right)}{\left( 1-{{u}^{2}} \right)\sqrt{s-\left| t \right|-4M_{\mu }^{2}}}. \\ 
\end{split}
\end{equation}
Покажемо, що величина ${{z}_{1}}$ є більшою за 1, а  ${{z}_{2}}$ є меншою за (-1) і таким чином знаменники в інтегралі \eqref{integririvanie_po_fi_1} не дорівнюють нулю всюди на проміжку інтегрування. 
Розглянемо різницю
\begin{equation}\label{riznicaz1minus1}
\begin{split}
& {{z}_{1}}-1=\frac{1}{\left( 1-{{u}^{2}} \right)\sqrt{s-\left| t \right|-4M_{\mu }^{2}}}\left( \sqrt{s}{{\left( u-\frac{m\left( t \right)}{\sqrt{s}}\left( 1+\frac{\left| t \right|}{4{{m}^{2}}\left( t \right)} \right) \right)}^{2}}+ \right. \\ 
& \left. +{{u}^{2}}\sqrt{s-\left| t \right|-4M_{\mu }^{2}}+\sqrt{s}\left( 1-\frac{{{m}^{2}}\left( t \right)}{s}{{\left( 1+\frac{\left| t \right|}{4{{m}^{2}}\left( t \right)} \right)}^{2}} \right)-\sqrt{s-\left| t \right|-4M_{\mu }^{2}} \right). \\ 
\end{split}
\end{equation}
Збираючи коефіцієнти при ступенях змінної $u,$ чисельник цього виразу може бути переписаний в виді:
\begin{equation}\label{chiselnic-pou_z1minus1}
\begin{split}
&\sqrt s \left( {u - \frac{{m\left( t \right)}}{{\sqrt s }}\left( {1 + \frac{{\left| t \right|}}{{4m^2 \left( t \right)}}} \right)} \right)^2  + u^2 \sqrt {s - \left| t \right| - 4M_\mu ^2 }  +  \\ 
&+ \sqrt s \left( {1 - \frac{{m^2 \left( t \right)}}{s}\left( {1 + \frac{{\left| t \right|}}{{4m^2 \left( t \right)}}} \right)^2 } \right) - \sqrt {s - \left| t \right| - 4M_\mu ^2 }  =  \\ 
&= \left( {\sqrt s  + \sqrt {s - \left| t \right| - 4M_\mu ^2 } } \right)u^2  - 2um\left( t \right)\left( {1 + \frac{{\left| t \right|}}{{4m^2 \left( t \right)}}} \right) +  \\ 
&+ \sqrt s  - \sqrt {s - \left| t \right| - 4M_\mu ^2 } . \\ 
\end{split}
\end{equation}
Дискриминант цього квадратичного по $u$ виразу дорівнює
\begin{equation}\label{Discriminant1}
D = 4\left( {\left( {M_G^2  + \frac{{\left| t \right|}}{4}} \right)\left( {1 + \frac{{\frac{{\left| t \right|}}{4}}}{{M_G^2  + \frac{{\left| t \right|}}{4}}}} \right)^2  - \left( {\left| t \right| + 4M_\mu ^2 } \right)} \right).
\end{equation}
Цей дискриминант може бути оцінений наступним чином:
\begin{equation}\label{ocinca_discriminantu1}
D < 4\left( {\left( {M_G^2  + \frac{{\left| t \right|}}{4}} \right)2^2  - \left( {\left| t \right| + 4M_\mu ^2 } \right)} \right) = 16\left( {M_G^2  - M_\mu ^2 } \right).
\end{equation}
Приймаючи що $ M_G  < M_\mu$ дістаємо висновку що дискримінант є від'ємний і чисельник виразу \eqref{riznicaz1minus1} не має нулів на дійсній вісі. Окрім того, як видно з \eqref{chiselnic-pou_z1minus1} при $ u=0 $ цей чисельник приймає значення 
$ \sqrt s  - \sqrt {s - \left| t \right| - 4M_\mu ^2 }, $ яке є додатним числом. Звідси маємо, що для будь-якого $ u $ різниця $z_1  - 1$
є додатною і $z_1  > 1.$ 

Розглянемо тепер різницю
\begin{equation}\label{z2miusminus1}
\begin{split}
&z_2  - \left( { - 1} \right) =  \\ 
&=  - \sqrt s \frac{{\left( {u + \frac{{m\left( t \right)}}{{\sqrt s }}\left( {1 + \frac{{\left| t \right|}}{{4m^2 \left( t \right)}}} \right)} \right)^2  + 1 - \frac{{m^2 \left( t \right)}}{s}\left( {1 + \frac{{\left| t \right|}}{{4m^2 \left( t \right)}}} \right)^2 }}{{\left( {1 - u^2 } \right)\sqrt {s - \left| t \right| - 4M_\mu ^2 } }} + 1 \\ 
\end{split}
\end{equation}

Цей вираз може бути перетворений таким чином:
\begin{equation}\label{z2miusminus1_preobr}
\begin{split}
&z_2  - \left( { - 1} \right) =  \\ 
&=  - \frac{{\left( {\sqrt s  + \sqrt {s - \left| t \right| - 4M_\mu ^2 } } \right)u^2  + 2um\left( t \right)\left( {1 + \frac{{\left| t \right|}}{{4m^2 \left( t \right)}}} \right) + \left( {\sqrt s  - \sqrt {s - \left| t \right| - 4M_\mu ^2 } } \right)}}{{\left( {1 - u^2 } \right)\sqrt {s - \left| t \right| - 4M_\mu ^2 } }} .\\ 
\end{split}
\end{equation}
Дискримінант чисельника цього виразу співпадає з \eqref{Discriminant1} і, тому є також від'ємним. Оскільки при $ u=0 $  вираз \eqref{z2miusminus1_preobr} приймає від'ємне значення, то й маємо що $ z_2  < \left( { - 1} \right). $ Отже в інтегралі \eqref{integririvanie_po_fi_1} знаменник ніде в області інтегрування не дорівнює нулю. 

Для подальшого розрахунку зручно ввести позначення 
\begin{equation}\label{c_1c_2 }
\begin{split}
c_1  = \frac{{\sqrt s \left( {1 + u^2 } \right) - 2m\left( t \right)u\left( {1 + \frac{{\left| t \right|}}{{4m^2 \left( t \right)}}} \right)}}{{\sqrt {s - \left| t \right| - 4M_\mu ^2 } }}, \\ 
c_2  =  - \frac{{\sqrt s \left( {1 + u^2 } \right) + 2m\left( t \right)u\left( {1 + \frac{{\left| t \right|}}{{4m^2 \left( t \right)}}} \right)}}{{\sqrt {s - \left| t \right| - 4M_\mu ^2 } }}. \\ 
\end{split}
\end{equation}
Тоді для \eqref{z1iz2} маємо вирази:
\begin{equation}\label{z_1z_2c_1c_2}
z_1  = \frac{{c_1 }}{{1 - u^2 }},z_2  = \frac{{c_2 }}{{1 - u^2 }}.
\end{equation} 

\begin{figure}
 	\centering
 	\includegraphics[scale=0.6]{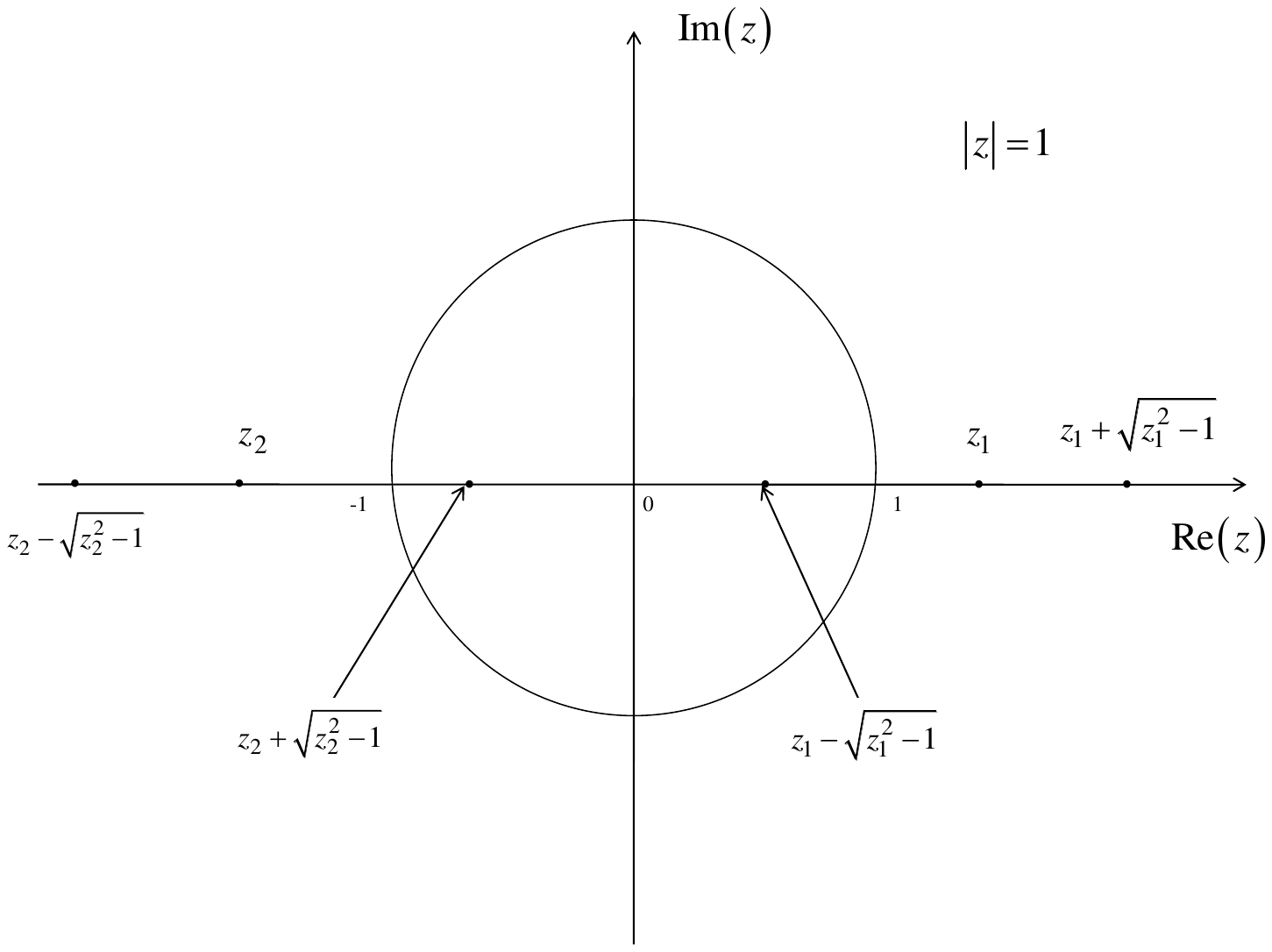}
 	\caption[]{Розташування полюсів підінтегрального виразу формули \eqref{Polusi_ed_kolo} з урахуванням виразу \eqref{Spivvidnohenna}}
 	\label{fig:roztahuvanna_polusiv_od_kolo}
 \end{figure}

Інтеграл по $  \phi  $ в \eqref{integririvanie_po_fi_1} може бути записаний в виді
\begin{equation}\label{Peretvorenna_do_conturnogo_integralu1}
\begin{split}
 &\int\limits_0^{2\pi } {d\varphi } \frac{1}{{\left( {\cos \left( \varphi  \right) - z_1 } \right)}}\frac{1}{{\left( { \cos \left( \varphi  \right) - z_2 } \right)}} =  \\ 
& = \int\limits_0^{2\pi } {d\varphi } \frac{1}{{\left( {\frac{{\exp \left( {i\varphi } \right) + \exp \left( { - i\varphi } \right)}}{2} - z_1 } \right)}}\frac{1}{{\left( {\frac{{\exp \left( {i\varphi } \right) + \exp \left( { - i\varphi } \right)}}{2} - z_2 } \right)}} \\ 
\end{split}
\end{equation}
Після перетворень цей вираз може бути переписаний таким чином:
\begin{equation}\label{Peretvorenna_do_conturnogo_integralu2}
\begin{split}
&\int\limits_0^{2\pi } {d\varphi } \frac{1}{{\left( {\cos \left( \varphi  \right) - z_1 } \right)}}\frac{1}{{\left( { + \cos \left( \varphi  \right) - z_2 } \right)}} =  \\ 
&=  - 4i\int\limits_0^{2\pi } {i\exp \left( {i\varphi } \right)d\varphi } \frac{1}{{\left( {\exp \left( {i\varphi } \right)} \right)^2  + 1 - 2z_1 \exp \left( {i\varphi } \right)}}\frac{{\exp \left( {i\varphi } \right)}}{{{\left( {\exp \left( {i\varphi } \right)} \right)^2  + 1 - 2z_2 \exp \left( {i\varphi } \right)}}} \\ 
\end{split}
\end{equation}
  Заміною змінних $ {\exp \left( {i\varphi } \right) = z} $ цей інтеграл може бути зведений до інтегралу по одиничному колу в комплексній площині, до якого може бути застосована теорема про лишки
  \begin{equation}\label{Peretvorenna_do_conturnogo_integralu}
  \begin{split}
&  \int\limits_0^{2\pi } {d\varphi } \frac{1}{{\left( {\cos \left( \varphi  \right) - z_1 } \right)}}\frac{1}{{\left( {\cos \left( \varphi  \right) - z_2 } \right)}} =  \\ 
&  =  - 4i\oint\limits_{\left| z \right| = 1} {dz} \frac{z}{{\left( {z^2  - 2z_1 z + 1} \right)\left( {z^2  - 2z_2 z + 1} \right)}}. \\ 
   \end{split}
  \end{equation}
 Для того, щоб виділити полюси підінтегрального виразу перепишемо його в виді
 \begin{equation}\label{Polusi_ed_kolo}
 \begin{split}
&\int\limits_0^{2\pi } {d\varphi } \frac{1}{{\left( {\cos \left( \varphi  \right) - z_1 } \right)}}\frac{1}{{\left( {\cos \left( \varphi  \right) - z_2 } \right)}} =  \\ 
&=  - 4i\oint\limits_{\left| z \right| = 1} {zdz} \left( {\frac{1}{{z - \left( {z_1  + \sqrt {z_1^2  - 1} } \right)}}\frac{1}{{z - \left( {z_1  - \sqrt {z_1^2  - 1} } \right)}}} \right. \times  \\ 
&\left. { \times \frac{1}{{z - \left( {z_2  + \sqrt {z_2^2  - 1} } \right)}}\frac{1}{{z - \left( {z_2  - \sqrt {z_2^2  - 1} } \right)}}} \right). \\ 
 \end{split}
 \end{equation} 
 З урахуванням співвідношень $z_1  > 1,z_2  < \left( { - 1} \right),$ отримаємо розташування полюсів по відношенню до контуру інтегрування, схематично показане на рис.\ref{fig:roztahuvanna_polusiv_od_kolo}.
 
При аналізі розташування полюсів ми скористалися співвідношеннями
\begin{equation}\label{Spivvidnohenna}
z_1  - \sqrt {z_1^2  - 1}  = \frac{1}{{z_1  + \sqrt {z_1^2  - 1} }},z_2  + \sqrt {z_2^2  - 1}  = \frac{1}{{z_2  - \sqrt {z_2^2  - 1} }}.
\end{equation}
Оскільки знаменники в обох виразах за модулем більші за одиницю, відповідні полюси мають модулі менші за одиницю і, таким чином потрапляють всередину одиничного кола, тобто всередину контуру, по якому відбувається інтегрування.
Застосування теореми про лишки призводить до результату
\begin{equation}\label{Rezultat_integruvanna_po_fi}
\begin{split}
&\int\limits_0^{2\pi } {d\varphi } \frac{1}{{\left( {\cos \left( \varphi  \right) - z_1 } \right)}}\frac{1}{{\left( { + \cos \left( \varphi  \right) - z_2 } \right)}} =  \\ 
&= 4\pi \left( {\frac{{z_2  + \sqrt {z_2^2  - 1} }}{{\sqrt {z_2^2  - 1} \left( {z_2  + \sqrt {z_2^2  - 1}  - \left( {z_1  + \sqrt {z_1^2  - 1} } \right)} \right)\left( {z_2  + \sqrt {z_2^2  - 1}  - \left( {z_1  - \sqrt {z_1^2  - 1} } \right)} \right)}} - } \right. \\ 
&\left. { - \frac{{z_1  - \sqrt {z_1^2  - 1} }}{{\sqrt {z_1^2  - 1} \left( {\left( {z_1  - \sqrt {z_1^2  - 1} } \right) - \left( {z_2  + \sqrt {z_2^2  - 1} } \right)} \right)\left( {\left( {z_1  - \sqrt {z_1^2  - 1} } \right) - \left( {z_2  - \sqrt {z_2^2  - 1} } \right)} \right)}}} \right). \\ 
\end{split}
\end{equation}
Використовуючи введені раніше позначення \eqref{c_1c_2 }, цей результат зручно переписати в виді 
\begin{equation}\label{integral_cerez_c}
\begin{split}
&\int\limits_0^{2\pi } {d\varphi } \frac{1}{{\left( {\cos \left( \varphi  \right) - z_1 } \right)}}\frac{1}{{\left( { + \cos \left( \varphi  \right) - z_2 } \right)}} =  \\ 
&= 4\pi \left( {1 - u^2 } \right)^2 \left( {\frac{1}{{\sqrt {\left( {c_2 } \right)^2  - \left( {1 - u^2 } \right)^2 } }}\frac{1}{{\left( {1 - \frac{{c_2  - \sqrt {\left( {c_2 } \right)^2  - \left( {1 - u^2 } \right)^2 } }}{{c_1  + \sqrt {\left( {c_1 } \right)^2  - \left( {1 - u^2 } \right)^2 } }}} \right)}} \times } \right. \\ 
&\times \frac{1}{{\left( {c_2  + \sqrt {\left( {c_2 } \right)^2  - \left( {1 - u^2 } \right)^2 }  - \left( {c_1  + \sqrt {\left( {c_1 } \right)^2  - \left( {1 - u^2 } \right)^2 } } \right)} \right)}} -  \\ 
&- \frac{1}{{\sqrt {\left( {c_1 } \right)^2  - \left( {1 - u^2 } \right)^2 } }}\frac{1}{{1 - \frac{{c_1  + \sqrt {\left( {c_1 } \right)^2  - \left( {1 - u^2 } \right)^2 } }}{{c_2  - \sqrt {\left( {c_2 } \right)^2  - \left( {1 - u^2 } \right)^2 } }}}} \times  \\ 
&\left. { \times \frac{1}{{\left( {\left( {c_1  - \sqrt {\left( {c_1 } \right)^2  - \left( {1 - u^2 } \right)^2 } } \right) - \left( {c_2  - \sqrt {\left( {c_2 } \right)^2  - \left( {1 - u^2 } \right)^2 } } \right)} \right)}}} \right). \\ 
\end{split}
\end{equation}
В результаті внесок підмножини 2 може бути зведений до одновимірного несингулярного інтегралу, який може бути розрахований, наприклад, методом трапецій:
\begin{equation}\label{vnesoc_mnogini2}
\begin{split}
&A_1^{\left\{ 2 \right\}}  = 16\pi G^4 \frac{1}{{m\left( t \right)\left( {s - \left| t \right| - 4M_\mu ^2 } \right)}} \times  \\ 
&\times \int\limits_0^1 {du\frac{{\left( {1 - u^2 } \right)^2 }}{{\left( {1 + u^2 } \right)}}} \left( {\frac{1}{{\sqrt {\left( {c_2 } \right)^2  - \left( {1 - u^2 } \right)^2 } }}} \right. \times  \\ 
&\times \frac{1}{{\left( {c_2  + \sqrt {\left( {c_2 } \right)^2  - \left( {1 - u^2 } \right)^2 }  - \left( {c_1  + \sqrt {\left( {c_1 } \right)^2  - \left( {1 - u^2 } \right)^2 } } \right)} \right)}} \times  \\ 
&\times \frac{1}{{\left( {1 - \frac{{c_2  - \sqrt {\left( {c_2 } \right)^2  - \left( {1 - u^2 } \right)^2 } }}{{c_1  + \sqrt {\left( {c_1 } \right)^2  - \left( {1 - u^2 } \right)^2 } }}} \right)}} -  \\ 
&- \frac{1}{{1 - \frac{{c_1  + \sqrt {\left( {c_1 } \right)^2  - \left( {1 - u^2 } \right)^2 } }}{{c_2  - \sqrt {\left( {c_2 } \right)^2  - \left( {1 - u^2 } \right)^2 } }}}}\frac{1}{{\sqrt {\left( {c_1 } \right)^2  - \left( {1 - u^2 } \right)^2 } }} \\ 
&\left. { \times \frac{1}{{\left( {\left( {c_1  - \sqrt {\left( {c_1 } \right)^2  - \left( {1 - u^2 } \right)^2 } } \right) - \left( {c_2  - \sqrt {\left( {c_2 } \right)^2  - \left( {1 - u^2 } \right)^2 } } \right)} \right)}}} \right). \\ 
\end{split}
\end{equation}

\section{Результати розрахунків}
\begin{figure}
	\centering
	\includegraphics[scale=0.2]{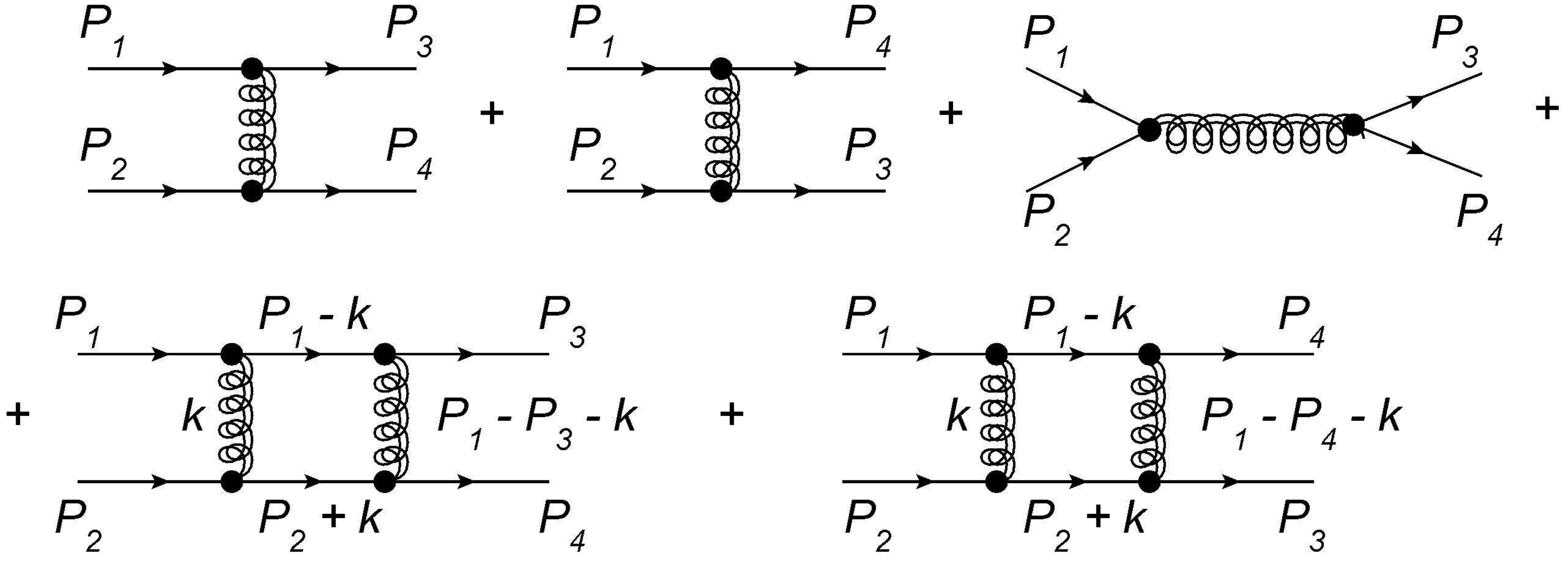}
	\caption[]{Сума діаграм пружного розсіяння мезонів в розглянутій моделі}
	\label{fig:suma_diagram_modeli}
\end{figure}

\begin{figure}
	\centering
	\includegraphics[scale=1.0]{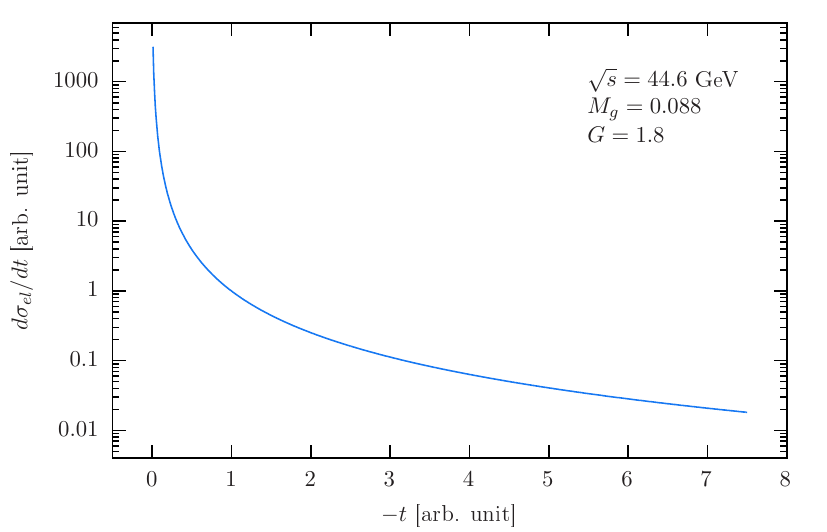}
	\caption[]{Результати розрахунку диференційного перерізу пружного мезон-мезонного розсіяння, розрахованого по описаній в цій роботі моделі. По обох вісях відкладено безрозмірні величини, $ t $ обезрозмірено на квадрат маси $\pi^0- $ мезону, а $\frac{d{{\sigma }_{el}}}{dt}$ на обернений квадрат маси $\pi^0- $ мезону.}
	\label{fig:dif_sechenie}
\end{figure}

Наведені раніше розрахунки дозволяють сформулювати модель пружного розсіяння мезонів, яка враховує суму діаграм, наведених на рис.\ref{fig:suma_diagram_modeli}. Тобто до діаграм, розглянутих вище ми додаємо діаграми, які від них відрізняються заміною між собою чотириімпульсів ${{P}_{3}}$ і ${{P}_{4}}.$ Окрім того, як видно з рис.\ref{fig:suma_diagram_modeli} додана сума безпетльових діаграм нижчого порядку. При цьому діаграми різного порядку теорії збурень входять в суму з різними коефіцієнтами, що пов'язано із різними ступенями константи зв'язку, множників $\left( {-i}/{{{\left( 2\pi  \right)}^{4}}}\; \right)$ що зіставляються лініям віртуальних частинок і вагових коефіцієнтах діаграм. Оскільки з точки зору зіставлення з експериментом, нас цікавить лише якісний вид диференційного перерізу пружного розсіяння за квадратом переданого чотириімпульса, ми можемо розраховувати суму діаграм на рис.\ref{fig:suma_diagram_modeli} з точністю до постійного коефіцієнта. При проведенні розрахунків, ми прийняли, що роль постійного коефіцієнта грає множник перед діаграмами другого порядка. Тобто, цей коефіцієнт ми виносимо за дужки, так що коефіцієнти при діаграмах другого порядку становляться рівними одиниці. Добуток всіх множників при діаграмах четвертого порядку позначатимемо $ L. $
Отже маємо модель з двома підгінними параметрами $ M_G $ і $ L. $ Всі величини розглядаються в безрозмірному виді. Обезрозмірювання проводилося на масу  $\pi^0- $ мезону, яка дорівнює $\approx 0.135$ГеВ. На Рис. \ref{fig:dif_sechenie} наведені результати розрахунку диференційного перерізу пружного розсіяння за допомогою описаної в роботі моделі з урахуванням діаграм на Рис. \ref{fig:suma_diagram_modeli}. Ці розрахунки проведені при таких величинах параметрів $\sqrt{s}=22.5$ ГеВ,$ L = 0.09 $,$ M_G=0.05 $. Нажаль, експериментальних даних з мезон-мезонного розсіяння, з якими можна було б порівняти отримані результати знайти не вдалося.

В цій роботі ми розглянули модель взаємодіючих скалярних полів, як найбільш просту для того, щоб сконцентруватися на розробці методу розрахунку петльової діаграми. Запропонований в цій роботі метод можна далі застосовувати до розрахунку петльових амплітуд розсіяння частинок із ненульовим спіном, однак при цьому ми очікуємо прояв механізмів утворення зростаючих із $\left| t \right|$, пов’язаних із суто спіновими ефектами.

\bibliographystyle{unsrtnat}
\bibliography{references-utf8} 

\begin{thebibliography}{20}
\providecommand{\natexlab}[1]{#1}
\providecommand{\url}[1]{\texttt{#1}}
\expandafter\ifx\csname urlstyle\endcsname\relax
  \providecommand{\doi}[1]{doi: #1}\else
  \providecommand{\doi}{doi: \begingroup \urlstyle{rm}\Url}\fi

\bibitem[Feynman(1949)]{PhysRev.76.769}
R.~P. Feynman.
\newblock Space-time approach to quantum electrodynamics.
\newblock \emph{Phys. Rev.}, 76:\penalty0 769--789, Sep 1949.
\newblock \doi{10.1103/PhysRev.76.769}.

\bibitem[Pancheri and Srivastava(2017)]{Pancheri2017}
Giulia Pancheri and Yogendra~N. Srivastava.
\newblock Introduction to the physics of the total cross section at lhc.
\newblock \emph{The European Physical Journal C}, 77\penalty0 (3):\penalty0
  150, Mar 2017.
\newblock ISSN 1434-6052.
\newblock \doi{10.1140/epjc/s10052-016-4585-8}.

\bibitem[Дрёмин(2013)]{Dryomin:2013}
И.~М. Дрёмин.
\newblock Упругое расcеяние адронов.
\newblock \emph{Усп. физ. наук}, 183\penalty0 (1):\penalty0 3--32,
  2013.
\newblock \doi{10.3367/UFNr.0183.201301a.0003}.

\bibitem[Никитин and Розенталь(1980)]{book:1030746}
Ю.~П. Никитин and И.~Л. Розенталь.
\newblock \emph{Ядерная физика высоких энергий}.
\newblock Атомиздат, 1980.

\bibitem[Gribov(1968)]{Gribov:1968fc}
V.~N. Gribov.
\newblock {A REGGEON DIAGRAM TECHNIQUE}.
\newblock \emph{Sov. Phys. JETP}, 26:\penalty0 414--422, 1968.
\newblock [Zh. Eksp. Teor. Fiz.53,654(1967)].

\bibitem[Baker and Ter-Martirosyan(1976)]{BAKER19761}
M.~Baker and K.A. Ter-Martirosyan.
\newblock Gribov's reggeon calculus: Its physical basis and implications.
\newblock \emph{Physics Reports}, 28\penalty0 (1):\penalty0 1 -- 143, 1976.
\newblock ISSN 0370-1573.
\newblock \doi{10.1016/0370-1573(76)90002-8}.

\bibitem[Кайдалов(2003)]{Kaidalov:2003}
А.~Б. Кайдалов.
\newblock Особенность Померанчука и
  взаимодействия адронов при высоких
  энергиях.
\newblock \emph{Успехи физических наук}, 173\penalty0
  (11):\penalty0 1153--1170, 2003.
\newblock \doi{10.3367/UFNr.0173.200311a.1153}.

\bibitem[Bourrely et~al.(2003)Bourrely, Soffer, and Wu]{Bourrely:2002wr}
Claude Bourrely, Jacques Soffer, and Tai~Tsun Wu.
\newblock {Impact picture phenomenology for $\pi^+$ -p, $K^+$ -p and $p p$,
  $\bar{p} p$ elastic scattering at high-energies}.
\newblock \emph{Eur.Phys.J.}, C28:\penalty0 97--105, 2003.
\newblock \doi{10.1140/epjc/s2003-01159-7}.

\bibitem[Fadin et~al.(1975)Fadin, Kuraev, and Lipatov]{FADIN197550}
V.S. Fadin, E.A. Kuraev, and L.N. Lipatov.
\newblock On the pomeranchuk singularity in asymptotically free theories.
\newblock \emph{Physics Letters B}, 60\penalty0 (1):\penalty0 50 -- 52, 1975.
\newblock ISSN 0370-2693.
\newblock \doi{10.1016/0370-2693(75)90524-9}.

\bibitem[Липатов(2008)]{Lipatov:2008}
Л.~Н. Липатов.
\newblock Бьёркеновская и реджевская
  асимптотики амплитуд рассеяния в
  квантовой хромодинамике и
  суперсимметричных калибровочных моделях.
\newblock \emph{Усп. физ. наук}, 178\penalty0 (6):\penalty0 663--668,
  2008.
\newblock \doi{10.3367/UFNr.0178.200806m.0663}.

\bibitem[Кураев et~al.(1977)Кураев, Липатов, and
  Фадин]{Kuraev:1977fs}
Э.~A. Кураев, Л.~Н. Липатов, and В.С. Фадин.
\newblock {Особенность Померанчука в
  неабелевых калибровочных теориях}.
\newblock \emph{ЖЭТФ}, 72:\penalty0 377--389, 1977.

\bibitem[Шарф et~al.(2011)Шарф, Тіхонов, Сохранний,
  Яткін, Делієргієв, et~al.]{Sharf:2011ufj}
І.~В. Шарф, А.~В. Тіхонов, Г.~О. Сохранний, К.~В.
  Яткін, M.~A. Делієргієв, et~al.
\newblock {Метод Лапласа для опису непружного
  розсіяння адронів і нові механізми
  зростання перерізів}.
\newblock \emph{УФЖ}, 56:\penalty0 1151--1164, 2011.

\bibitem[Sharph et~al.(2012)Sharph, Tykhonov, Sokhrannyi, Deliyergiyev,
  Podolyan, et~al.]{Sharph:2011wm}
Igor Sharph, Andrii Tykhonov, Grygorii Sokhrannyi, Maksym Deliyergiyev, Natalia
  Podolyan, et~al.
\newblock {On the Role of Longitudinal Momenta in High Energy Hadron-Hadron
  Scattering}.
\newblock \emph{Central Eur.J.Phys.}, 10:\penalty0 858--887, 2012.
\newblock \doi{10.2478/s11534-012-0056-5}.

\bibitem[Easther et~al.(2000)Easther, Guralnik, and Hahn]{PhysRevD.61.125001}
Richard Easther, Gerald Guralnik, and Stephen Hahn.
\newblock Fast evaluation of feynman diagrams.
\newblock \emph{Phys. Rev. D}, 61:\penalty0 125001, May 2000.
\newblock \doi{10.1103/PhysRevD.61.125001}.

\bibitem[HEINRICH(2008)]{doi:10.1142/S0217751X08040263}
GUDRUN HEINRICH.
\newblock Sector decomposition.
\newblock \emph{International Journal of Modern Physics A}, 23\penalty0
  (10):\penalty0 1457--1486, 2008.
\newblock \doi{10.1142/S0217751X08040263}.

\bibitem[Li et~al.(2016)Li, Wang, Yan, and Zhao]{Li_2016}
Zhao Li, Jian Wang, Qi-Shu Yan, and Xiaoran Zhao.
\newblock Efficient numerical evaluation of feynman integrals.
\newblock \emph{Chinese Physics C}, 40\penalty0 (3):\penalty0 033103, mar 2016.
\newblock \doi{10.1088/1674-1137/40/3/033103}.

\bibitem[Де~Брёйн(1961)]{book:5160}
Н.Г. Де~Брёйн.
\newblock \emph{Асимптотические методы в анализе}.
\newblock 1961.

\bibitem[Sharf et~al.(2012)Sharf, Merkotan, Podolyan, Ptashynskyy, Tykhonov,
  et~al.]{Sharf:2012vy}
I.V. Sharf, K.K. Merkotan, N.A. Podolyan, D.A. Ptashynskyy, A.V. Tykhonov,
  et~al.
\newblock {Gluon Loops in the Inelastic Processes in QCD}.
\newblock 2012.
\newblock arXiv:1210.3490.

\bibitem[Chudak et~al.(2016)Chudak, Deliyergiyev, Merkotan, Potiienko,
  Ptashynskyi, Shabatura, Sokhrannyi, Tykhonov, Volkotrub, Sharph, and
  Rusov]{Chudak:2016}
N.~Chudak, M.~Deliyergiyev, K.~Merkotan, O.~Potiienko, D.~Ptashynskyi,
  Y~Shabatura, G.~Sokhrannyi, A.~Tykhonov, Y.~Volkotrub, I.~Sharph, and
  V.~Rusov.
\newblock Multi-particle quantum fields.
\newblock \emph{Physics Journal}, 2\penalty0 (3):\penalty0 181--195, 2016.

\bibitem[O.~Chudak et~al.(2019)O.~Chudak, K.~Merkotan, A.~Ptashynskiy,
  S.~Potiienko, Sharph, and I.~Bregid]{articleJFS}
N~O.~Chudak, K~K.~Merkotan, D~A.~Ptashynskiy, O~S.~Potiienko, I~Sharph, and
  V~I.~Bregid.
\newblock The calculation of the differential cross section of hadron elastic
  scattering by transferred four-momentum within the perturbation theory.
\newblock \emph{Journal of Physical Studies}, 23, 01 2019.
\newblock \doi{10.30970/jps.23.1101}.

\end{thebibliography}
 

\end{document}